\documentclass[nofootinbib,showkeys,floatfix,times]{revtex4-2}
\usepackage{bm,morefloats}
\usepackage{dcolumn}
\usepackage[latin1]{inputenc}
\usepackage[T2A,T1]{fontenc} % load the desired encodings
\usepackage{fonttable}
\usepackage{graphicx}
\usepackage{balance}
\usepackage[latin1]{inputenc}
\usepackage[T1]{fontenc}
        \usepackage{cellspace}
\usepackage{latexsym,color}
\usepackage{centernot}
\usepackage{mathtools}
\usepackage{stmaryrd}
\usepackage{amssymb}
\usepackage{natbib}
\usepackage{amsmath,amssymb,amsthm,mathrsfs,amsfonts,dsfont}

\usepackage{color}

\usepackage{rotating}
\usepackage[dvipsnames]{xcolor}
\definecolor{LinkBlue}{RGB}{6,69,173}
\definecolor{DarkBlue}{RGB}{11,0,128}
\definecolor{red}{rgb}{1,0.,0.}
\usepackage[colorlinks=true,linkcolor=LinkBlue,urlcolor=magenta,
	citecolor=green,hyperfootnotes=true,urlcolor=red]{hyperref}

\usepackage{enumitem}
\count\footins = 1000

\newcommand{\mnras}{MNRAS}

\newcommand{\prdd}{Phys. Rev. \textbf{D}}

\begin{document}

\title{Inversion  points of the  accretion flows onto    super-spinning Kerr attractors}
\author{D. Pugliese\&Z. Stuchl\'{\i}k}
\email{daniela.pugliese@physics.slu.cz}
\affiliation{
%den\v{e}k
Research Centre for Theoretical Physics and Astrophysics, Institute of Physics,
  Silesian University in Opava,
 Bezru\v{c}ovo n\'{a}m\v{e}st\'{i} 13, CZ-74601 Opava, Czech Republic
%5
}

\begin{abstract}We study the accretion flows  towards a  central Kerr super-spinning attractor, discussing
the formation   of  the  flow inversion points, defined by condition $u^{\phi}=0$ on the particles flow axial  velocity.
We locate two   closed  surfaces, defining    \emph{inversion coronas} (spherical shells), surrounding the central attractor. The coronas analysis  highlights observational aspects  distinguishing the central attractors and  providing  indications  on their  spin and     the  orbiting  fluids.
 The inversion corona is  a closed  region, generally of small extension and thickness, which is  for the counter-rotating flows of the order of $\lesssim 1.4 M$ (central attractor mass)  on the vertical rotational  axis.
 There are  no co-rotating   inversion points  (from co-rotating flows). The   results point to  strong    signatures of the  Kerr super-spinars, provided  in both accretion  and jet flows.
With   very narrow thickness, and varying  little with the   fluid initial conditions and the emission process details, inversion coronas can  have remarkable  observational significance for  primordial Kerr super-spinars predicted   by string theory. The  corona region closest to the central attractor is the most observably recognizable and active part, distinguishing black holes solutions from super-spinars.  Our analysis expounds the    Lense--Thirring  effects and repulsive gravity effects in the super-spinning ergoregions.
\end{abstract}
\date{\today}

\maketitle

\def\be{\begin{equation}}
\def\ee{\end{equation}}
\def\bea{\begin{eqnarray}}
\def\eea{\end{eqnarray}}
\newcommand{\tb}[1]{\textbf{\texttt{#1}}}
\newcommand{\actaa}{Acta Astronomica}
\newcommand{\laa}{\mathcal{L}}
\newcommand{\ba}{\mathcal{B}}
\newcommand{\Sie}{\mathcal{S}}
\newcommand{\Mie}{\mathcal{M}}
\newcommand{\La}{\mathcal{L}}
\newcommand{\Em}{\mathcal{E}}

\newcommand{\ms}{\mathrm{ms}}
\newcommand{\mb}{\mathrm{mb}}

\newcommand{\rtb}[1]{\textcolor[rgb]{1.00,0.00,0.00}{\tb{#1}}}
\newcommand{\gtb}[1]{\textcolor[rgb]{0.0,0.52,0.0}{#1}}
\newcommand{\ptb}[1]{\textcolor[rgb]{0.77,0.04,0.95}{\tb{#1}}}
\newcommand{\btb}[1]{\textcolor[rgb]{0.00,0.00,1.00}{#1}}
\newcommand{\otb}[1]{\textcolor[rgb]{1.00,0.50,0.25}{\tb{#1}}}
\newcommand{\non}[1]{{\LARGE{\not}}{#1}}

\newcommand{\cc}{\mathrm{C}}

\newcommand{\il}{~}
\newcommand{\la}{\mathcal{A}}
  \newcommand{\Qa}{\mathcal{Q}}
\newcommand{\Sa}{\mathcal{\mathbf{S}}}
\newcommand{\Ta}{{\mbox{\scriptsize  \textbf{\textsf{T}}}}}
\newcommand{\Ca}{\mathcal{\mathbf{C}}}

\section{Introduction}
This work explores astrophysical tracers distinguishing Kerr super-spinning solutions, with dimensionless spin $a>1$  from Kerr black hole ({BH}) solutions with $a<1$.

We focus on  the (accretion) flows  towards   a central super-spinning Kerr attractor, exploring
 the  formation of azimuthal velocity inversion points in   the accretion flows, defined  by the  condition $u^{\phi}=0$ on the  particles  axial velocity. The  analysis follows paper \cite{published}, where the case of   Kerr {BH} attractors  has been considered and \cite{submitted}, and where the case of   Kerr Naked singularity {(NS)}   has been detailed.
The inversion points in the  super-spinar  geometries are related to a combination of  the  repulsive gravity effects and  the frame dragging  effects, typical of  the ergoregion.

  The possibility to distinguish  between {BHs} and super-spinars  is still  a major challenge of the present day Astrophysics.
A  conclusive proof of  the  {C}osmic
{C}ensorship {h}ypothesis   ({CCh})   is  still lacking at present.,  on the other hand, super-spinars may appear in a variety of contexts, implying, in some cases, effects of quantum gravity or string theory \cite{Horava,z014,z01}.
Super-spinars can appear for example  in the frame of string theory.
The region of causality violations and  ring---like Kerr singularity
defining the  Kerr naked-singularity ({NS})    can be  removed and substituted by a spacetime  region represented  by a  string  solution.
These "primordial" super-spinars would be distinguished by an   exterior  geometry of a Kerr {NS}.
Then,  Kerr {NSs}   could
exist, without contradicting with the {CCh}, as remnants of a primordial high-energy phase of the  early
Universe \cite{Horava}.

{
Kerr super-spinars  are devoid of  the  physical ring
singularity  and the  causality-violation region,  substituted by an interior regular solution.
(The  removed pathological causality violation region is determined  from the condition  $g_{\phi\phi}<0$, occurring at $r < 0$.).
 Consequently, the  minimal condition for the boundary surface of Kerr super-spinars  is $r(\theta)=R=0$ (in the Boyer--Lindquist coordinates) although,  in many analyses, the Kerr super-spinars are considered at
the boundary $R=0.1M$ \cite{z014,z01,z04,z03}.
As consequence of this, the Kerr super-spinars  do not  contradict the {CCh}  (on the other hand, {CCh} counterexamples, where  existence of spherically symmetric {NSs}  or  evidences
of a violation of the weak
{CCh} in five-dimensional spacetimes   are often debated \cite{Figueras}). The  region of causality violation (at $r<R$=constant)  would be  covered,  for example, by a stringy solution,  expected  to be  limited to the region
$r < R \leq M$  \cite{Horava,z014,z01,z04,z03}.
The observation of such Kerr super-spinars could  be  therefore  interpreted in the framework of the string theory.
Outside a Kerr super-spinar ($r>R$), the  Kerr {NS} geometry is assumed.
The   Kerr super-spinars surface
 (at $r=R>0$) has to be assumed, and in some studies  the solution of a one-way membrane (similarly to the {BH} horizon) is considered \cite{z04,z01,z014}
(the surfaces  have  to be provided  by the exact interior solution to be joined at the boundary to the standard Kerr {NS} geometry).
However, contrary to the {BHs}, there is no uniqueness theorem for   the (Kerr) {NSs} (and super-spinars).
Kerr super-spinars are considered  as a reinterpretation of {NSs} possibly modeling  quasars of the early evolution period
of the Universe.
In this context therefore observations of a hypothetical Kerr super-spinar could  be expected at high
redshift AGN and quasars\cite{2011CQGra..28o5017S}.
In  theories different from General Relativity, such as braneworlds.  the Kerr {BH} spin limit $a=1$ can be exceeded \cite{Aliev2005,Aliev2009,2016PhRvD..94h6006B,Stuchlik:2008fy,brane}.

For all these reasons,   the study  of  possible  astrophysical signatures   for
Kerr super-spinars   constitutes an extensive  and debated literature on a  huge variety
of astrophysical phenomena--see for example \cite{z01,z04,z05,z06,z09,z010,z011,z013,1980BAICz..31..129S,1981BAICz..32...68S,z014,z015,z017}.

In the present analysis, we  discuss  some relevant astrophysical
phenomena exposing distinctions between
Kerr {NSs} and {BHs}, related to   flows from   accretion tori and  jets in the super-spinar gravitational  field.
Discussing the differences emerging  from  this analysis, between the {BH} and {NS} scenarios,  we  provide  distinguishable physical characteristics of  jet  and accretion flows in  {NS} geometries,  { which possibly could  provide   an
observational signature of the Kerr {NSs}}.

Accretion has been studied  for  possible mechanisms  to   convert
{NSs} into   extreme {BH} states\footnote{There are strong indications that   the inverse process, consisting in
the creation of  a {NS} through a conversion process from an extreme {BH}, is forbidden \cite{bardeen}}.
It  has  been shown that, following  accretion from orbiting   disks, Kerr super-spinars  may be   efficiently converted to
extreme Kerr (or near extreme Kerr) {BHs}\cite{1980BAICz..31..129S,1981BAICz..32...68S}.
The conversion into a Kerr {BH}, following  accretion from counter-rotating  Keplerian discs,  could be
 very quick. (However, crucially {NSs} from a primordial era could survive to the era of high--redshift quasars or even in  later times
 $z\sim2$  for a sub Eddington accretion process)\footnote{
 The efficiency of the counter-rotating machine is  by orders smaller of the
co-rotating  accretion  disk  conversion efficiency.)
 In this respect, Kerr super-spinars  can  also be efficient accelerators for extremely high--energy collisions--\cite{z03,z016}.}.

 Kerr {NS} spacetimes are characterized by an articulated combination of  Lense-Thirring and the repulsive gravity effects\footnote{It should be stressed that here the gravitational repulsion is not related to the vacuum energy effect as in inflationary models, but only to special configurations of attractive gravitational fields as demonstrated by F. de Felice \cite{1974A&A....34...15D}.}, peculiar of the  {NS} ergoregion.
 In the inversion point analysis, the counter-rotating  flows and the  {NS}  ergoregion are a particular focus.
We here  distinguish counter-rotating and co-rotating  flows {from accretion disks (\emph{"accretion driven flows"}) or from  structures   constituted by open     funnels of matter along the central attractor rotational axis (these structures are well known to  be  associated to  accreting toroids, see for example \cite{KJA78,AJS78,Sadowski:2015jaa,Lasota:2015bii,Lyutikov(2009),Madau(1988),Sikora(1981),abrafra}, and are here called, for brevity, \emph{proto-jets})--("\emph{proto-jets driven flows}"). We also considered    the formation of double toroidal  structures}  with equal specific angular momentum $\ell$, and ZAMOS (zero angular momentum observers) tori (with fluid specific  angular momentum  $ \ell = 0 $), typical of  {NS} geometries.
Two main scenarios are therefore  addressed:  the accretion driven inversion points and  the proto-jets driven inversion points, where the flux  of particles and photons is related to  an open cusped  toroidal configuration with high magnitude of  specific fluid angular momentum and variously associated to jets emission \cite{abrafra,proto-jets,
ella-correlation,ringed,dsystem,open,long}.
This setup fixes particle trajectories initial data  and constants of motion.

Although the inversion points do not depend on the details of the accretion processes, or the precise location of the tori  inner edge, in our analysis the flow of  free-falling particles is  driven from an inner edge of the accretion torus or proto-jet (open, cusped configuration).
As for the {BH} case, analyzed in \cite{published}, the  flow inversion points  in the  {NS} spacetimes,  define a closed and regular surface at fixed $\ell$, \emph{inversion surface}, surrounding  the central attractor, being influenced by the range of location of the  torus inner edge or the proto-jet cusps.

\medskip

More in details,  article plan is as follows:

In Sec.\il(\ref{Sec:quaconsta}) we introduce the spacetime metric  and the constants of motion.  In Sec.\il(\ref{Sec:coro-contro}) we explore  further the notion of  co-rotation and counter-rotation in {NSs} spacetimes.
Spacetime geodesic structure and tori constraints are   addressed in Sec.\il(\ref{Sec:extended-geo-struc}).
In Sec.\il(\ref{Sec:flow-inversion-points}),
  we define the flow inversion points and detail general properties, as  inversion points limits and extremes, for co-rotating and counter-rotating  flows.
 In Sec.\il(\ref{Sec:all-to-rediid}) inversion points of the  flows are constrained according to the {NS} spin, the fluid specific angular momentum and  the particles energy.
Sec.\il(\ref{Sec:inversion-counter-rot}) focuses on the counter-rotating flows, discussing
inversion radius  $r_\Ta^{\pm}$ in Sec.\il(\ref{Sec:counter-rotating-inversion-NSflow}) and the
inversion plane $\sigma_\Ta$ in Sec.\il(\ref{Sec:inversionsigmacontro}).
Constants of motion and inversion points in counter-rotating flows are  the subject of Sec.\il(\ref{Sec:sever-amernothermed}).
Inversion points of the co-rotating  flows are studied  in Sec.\il(\ref{Sec:corot-inversion}). Finally we explore the limiting case of tori with $\ell=0$ in Sec.\il(\ref{Sec:inversionlequal0}). 
Sec.\il(\ref{Sec:discussion}) contains discussion and final remarks.
{In Appedix\il(\ref{Appendix:radii}) we  explicit
Kerr Naked singularities spacetime  circular geodesic  radii.
 We also introduced Table\il(\ref{Table:Notation}), as a look-up table we refer the reader to for reminding  of notation and definitions  used throughout this paper with links to associated sections, equations  or  figures.}
\section{The spacetime metric and constants of motion}\label{Sec:quaconsta}
The   Kerr  spacetime metric  reads
%r
%
\bea \label{alai}&& ds^2=-\left(1-\frac{2Mr}{\Sigma}\right)dt^2+\frac{\Sigma}{\Delta}dr^2+\Sigma
d\theta^2+\left[(r^2+a^2)+\frac{2M r a^2}{\Sigma}\sin^2\theta\right]\sin^2\theta
d\phi^2
-\frac{4rMa}{\Sigma} \sin^2\theta  dt d\phi,
\eea
in the Boyer-Lindquist (BL)  coordinates
\( \{t,r,\theta ,\phi \}\)\footnote{We adopt the
geometrical  units $c=1=G$, Latin indices run in $\{0,1,2,3\}$.  The radius $r$ has unit of
mass $[M]$, and the angular momentum  units of $[M]^2$, the velocities  $[u^t]=[u^r]=1$
and $[u^{\phi}]=[u^{\theta}]=[M]^{-1}$ with $[u^{\phi}/u^{t}]=[M]^{-1}$ and
$[u_{\phi}/u_{t}]=[M]$. For the seek of convenience, we always consider the
dimensionless  energy and an angular momentum per
unit of mass $[L]/[M]=[M]$.},
where
%
%\bea\label{Eq:delta}
$
\Delta\equiv a^2+r^2-2 rM\quad\mbox{and}\quad \Sigma\equiv a^2 cos^2\theta+r^2$.
%\eea
%
 The Kerr {NSs} have  $a>M$.
 Parameter  $a=J/M\geq0$ is the metric spin, where  total angular momentum is   $J$  and  the  gravitational mass parameter is $M$.
  A Kerr {BH} is defined  by the condition $a\in \,[\,0,M\,]$  with Killing horizons horizons $r_-\leq r_+$ where  $r_{\pm}\equiv M\pm\sqrt{M^2-a^2}$).
  The extreme Kerr {BH}  has dimensionless spin $a/M=1$ and the non-rotating   case $a=0$ is the   Schwarzschild {BH} solution.

In the following we will use the angular coordinate $ \sigma\equiv \sin^2\theta$.
The equatorial plane, $\sigma=1$, is a metric symmetry plane  and   the  equatorial  circular geodesics are confined on the equatorial  plane as a consequence of the metric tensor symmetry under reflection through  the plane $\theta=\pi/2$.
In this work we address the {NSs} case, focusing in particular on the Lense-Thirring effects induced  from the frame-dragging,  {particularly relevant in the  ergoregion, whose boundaries are  the  outer  and inner stationary  limits $r_{\epsilon}^\pm$ (ergosurfaces),   given by}
\bea\label{Eq:sigma-erg}
r_{\epsilon}^{\pm}\equiv M\pm\sqrt{M^2- a^2 (1-\sigma)}\quad \mbox{or, equivalently}\quad \sigma_{erg}\equiv \frac{(r-2M) r}{a^2}+1,
%&&\nonumber
\eea
respectively---Figs\il(\ref{Fig:Plotvinthallosrc12WW})--where  $r_\epsilon^-=0$  and
  $r_{\epsilon}^+=2M$  in the equatorial plane $\theta=\pi/2$ ($\sigma=1$),
and  $r_+<r_{\epsilon}^+$ on   $\theta\neq0$.

More precisely  there is
\bea\label{Eq:fioc-rumor}&&
g_{tt}>0\; \mbox{for}:a> M,\quad \sigma \in \, \,]\,\sigma_\epsilon, 1\,]\,,r\in \,]\,r_{\epsilon}^-,r_{\epsilon}^+\,[\, ,\quad
%\\
%&&\label{Eq:sigmaepsilonplus}
\mbox{where}\;
\sigma_\epsilon\equiv\frac{a^2-M^2}{a^2}.
\\&&\nonumber
g_{tt}<0\; \mbox{for}:a>M,
\left(\sigma\in \,[\,0,\sigma_\epsilon\,[\, , r\geq 0\right),\quad
\left(\sigma =\sigma_\epsilon,\quad
r\neq r_\epsilon^-\right),\quad\left(
\sigma\in \,]\,\sigma_\epsilon,1\,]\, ,\quad r\in \, [\,0,r_{\epsilon}^-\,[\, \cup \, ]\,r_\epsilon^+,+\infty\,[\, \right),
%\\\nonumber
%&&
\eea
(see Figs\il(\ref{Fig:Plotvinthallosrc12WW})). {NS}  poles  $\sigma=0$ are associated to $g_{tt}<0$ for $a\geq0$  and $r\geq0$.
\begin{figure}
      \includegraphics[width=8cm]{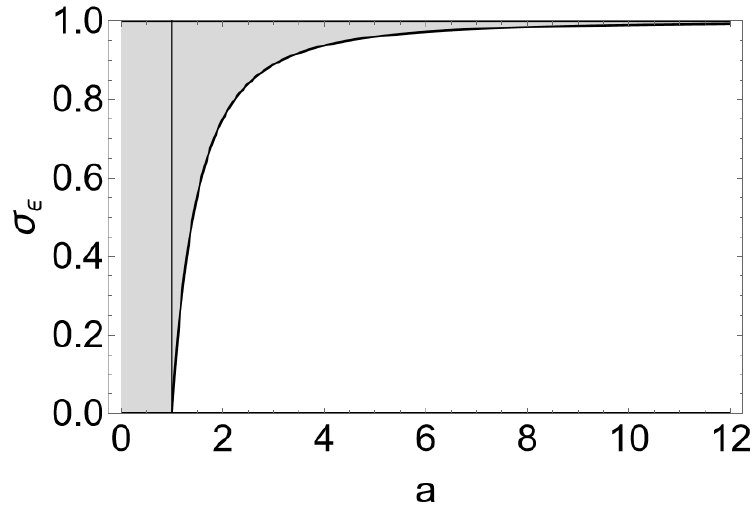}
          \includegraphics[width=8cm]{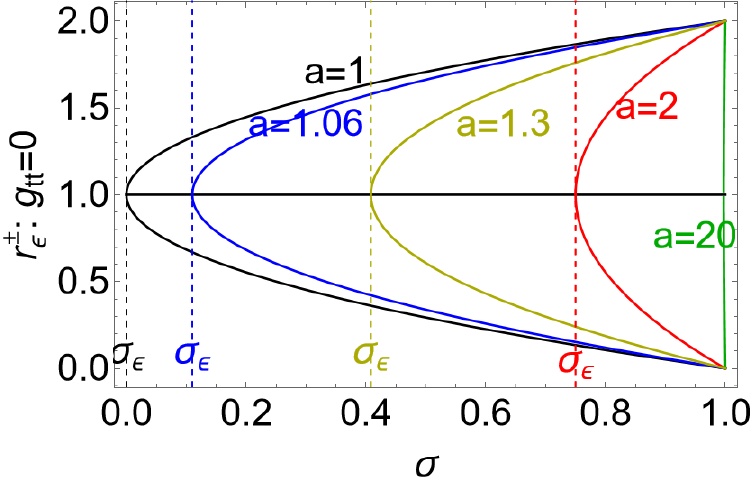}
            \caption{Left panel: Plane $\sigma_\epsilon$ in Eq.\il(\ref{Eq:fioc-rumor}) as function of the {NS}  spin $a>1$. The ergoregion is defined for $\sigma\in[\,\sigma_\epsilon,1\,]$. There is $\sigma\equiv \sin^2\theta\in[\,0,1\,]$, where $\sigma=1$ is the equatorial plane.    Radius $r=r_{\epsilon}^+=2$ is the outer-ergosurface on the equatorial plane. Right panel: the curves  (ergosurfaces)  $r_\epsilon^\pm$ such that  $g_{tt}(r_\epsilon^\pm)=0$ are plotted for selected spin, signed on the panel in different colors
on the plane $(r,\sigma)$. The  limiting values $\sigma_\epsilon$ for each spin are also shown as vertical colored dashed lines.
 There is $g_{tt}>0$ in the region bounded by the curves $r_\epsilon^\pm$.
 See analysis in Eq.\il(\ref{Eq:fioc-rumor}).   All quantities are dimensionless.} \label{Fig:Plotvinthallosrc12WW}
\end{figure}
The role of the limit $\sigma_\epsilon$  in the {NS} geometries is a main difference with the  {BH}
cases---see Figs\il(\ref{Fig:Plotvinthallosrc12WW}).
For  very fast super-spinning {NSs} $(a\gg M)$, there is $\sigma_\epsilon\to1$ and the range for the existence of the ergoregion narrows /at $a=M$ there is $\sigma_\epsilon=0)$--Figs\il(\ref{Fig:Plotvinthallosrc12WW}).
In the following, where appropriate,  to easy the reading of   complex expressions, we will  use  the   units with $M=1$ (where $r\rightarrow r/M$  and $a\rightarrow a/M$).

\medskip

\textbf{Constants of motion}

We consider  the  following  constant of geodesic  motions
\bea&&\label{Eq:EmLdef}
\Em=-(g_{t\phi} \dot{\phi}+g_{tt} \dot{t}),\quad \La=g_{\phi\phi} \dot{\phi}+g_{t\phi} \dot{t},\quad  g_{ab}u^a u^b=\kappa,
\eea
 defined  from   the Kerr geometry  rotational  Killing field   $\xi_{\phi}=\partial_{\phi}$,
     and  the Killing field  $\xi_{t}=\partial_{t}$
representing the stationarity of the  background,
with   $u^a\equiv\{ \dot{t},\dot{r},\dot{\theta},\dot{\phi}\}$ where
$\dot{q}$ indicates the derivative of any quantity $q$  with respect  the proper time or  a properly defined  affine parameter for the light--like orbits (for $\kappa=0$), and  $\kappa=(\pm1,0)$ is a normalization constant ($\kappa=-1$ for  test particles).

 The constant $\La$ in Eq.\il(\ref{Eq:EmLdef}) may be interpreted       as the axial component of the angular momentum  of a test    particle following
timelike geodesics and $\Em$  represents the total energy of the test particle
 related to the  radial infinity, as measured  by  a static observer at infinity.
The velocity components $(u^t,u^\phi)$ read
\bea\label{Eq:ufidottdotinconst}
u^t= \frac{g_{\phi\phi} \Em+ g_{t\phi} \La}{g_{t\phi}^2-g_{\phi\phi} g_{tt}},\quad u^\phi= -\frac{g_{t\phi} \Em+ g_{tt} \La}{g_{t\phi}^2-g_{\phi\phi} g_{tt}}.
\eea
We  introduce also  the specific  angular momentum $\ell$ and the relativistic angular velocity   $\Omega \equiv{u^\phi}/{u^{t}}$
 \bea&&\label{Eq:flo-adding}
\ell\equiv\frac{\La}{\Em}=-\frac{g_{\phi\phi}u^\phi  +g_{\phi t} u^t }{g_{tt} u^t +g_{\phi t} u^\phi} =-\frac{g_{t\phi}+g_{\phi\phi} \Omega }{g_{tt}+g_{t\phi} \Omega},\quad  \Omega(\ell)= -\frac{g_{t\phi}+ g_{tt} \ell}{g_{\phi\phi}+ g_{t\phi} \ell}.
\eea
%

%
 %\end{alertblock}
 {By considering definition of constants of geodesic motion $\La$ and $\ell$ in  Eq.\il(\ref{Eq:EmLdef}) and   Eq.\il(\ref{Eq:flo-adding}) we discuss in Sec.\il(\ref{Sec:coro-contro}) the notion of
 co-rotation and counter-rotation in the  Kerr {NS} spacetime.
 These definitions will constitute a primary template of analysis of the flows we investigate in this work and the inversion surfaces as (newly introduced) geometric properties of the Kerr NSs.
Finally, spacetime geodesic structure and tori constraints are   the focus of Sec.\il(\ref{Sec:extended-geo-struc}). In this section we will explore the circular geodesic  co-rotation and counter-rotation motion of the Kerr  NSs.
These radii constrain the circular test particle geodesics and   toroidal extended matter orbiting around the central spinning attractor as, for example, accretion disks. The flows we study are made  of matter  (test particles) and photons,   freely moving  towards the central attractor or from the central attractors (we address these different aspects  more closely also in Sec.\il(\ref{Sec:all-to-rediid})). The motion is  regulated by  the geodesic equations and can be parametrized by the constants of motion in Eq.\il(\ref{Eq:EmLdef}). We will use   the constant $\ell$, as it is   the  parameter defining the properties focused in this analysis.
It  is useful to classify the flows and   the characteristics of the accreting structures  which could  possibly engine the flows,  according to  the $\ell$ parameter ranges, as studied in Sec.\il(\ref{Sec:extended-geo-struc}), where we  discuss the flows   and accretion tori properties, providing constraints for our set-up. } %
\subsection{On co-rotation and counter-rotation in {NSs} spacetimes}\label{Sec:coro-contro}
{In this section we introduce some of the constraints and parameters  used in this analysis, discussing
 the specific angular momentum $\ell$ parametrizing  the flows, orbiting  tori and  inversion points. We shall examine  both co--rotating and counter---rotating motion with respect to the central attractor, investigating  their definitions properties in the Kerr NS spacetimes.}

The specific angular momentum $\ell=$constant parametrizes the  GRHD barotropic geometrically thick disks-- \cite{abrafra}.
These toroids are
constant pressure  surfaces, whose construction in the axes--symmetric spacetimes  is   based  on the application  of the von Zeipel theorem. The surfaces of constant angular velocity $\Omega$ and of constant specific angular momentum $\ell$ coincide. The toroids  rotation law, $\ell=\ell(\Omega)$, is therefore  independent of  the  details of the equation of state,  providing  the   integrability condition of   the Euler equation in the case of  barotropic fluids.  Consequently, in the geometrically thick disks,  the functional form of the angular
momentum and entropy distribution, during the evolution of dynamical processes, depends on the initial conditions of the system and not on
the details of the dissipative processes \cite{abrafra}.

We analyze  two families of  flows,  with fluid specific angular momentum $\ell=\ell^+<0$ (counter-rotating) and $\ell=\ell^-\lesseqgtr0$ (co-rotating and counter-rotating) respectively\footnote{{We also used notation  $\ell^\pm$ both for the function  $\ell^\pm$ and referring to the  constant values (and  constant of motions)  $\ell^\pm=$constant  characterizing  orbits, tori or inversion surfaces. Saying differently function $\ell^\pm$ of Eq.\il(\ref{Eq:pian-l-equa-l}) provides the (radial) distribution of the constant of motion $\ell$ for each orbit.  Figs\il(\ref{Fig:PlotdicsprifUaosc}) show the distribution
$\ell^\pm$ as function of $r$ for different spin $a$, for NSs and BHs.}}:
{
\bea\label{Eq:pian-l-equa-l}
\ell^{\mp}\equiv\frac{a^3\mp r^{3/2} \sqrt{\Delta^2}-a (4-3 r) r}{a^2-(r-2)^2 r},
\eea
\cite{abrafra,slany,pugtot}}. Note that  tori are governed by the distribution of specific angular momentum on the equatorial plane, therefore $\ell^\pm$  in Eqs\il(\ref{Eq:pian-l-equa-l}) are  independent of the $\sigma$-coordinate.
\begin{figure}
   \includegraphics[width=8cm]{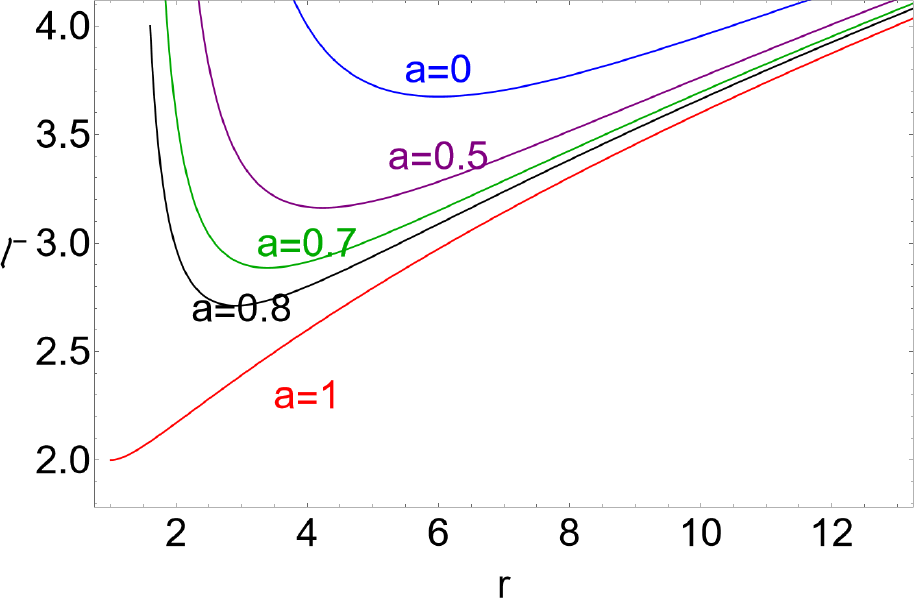}
   \includegraphics[width=8cm]{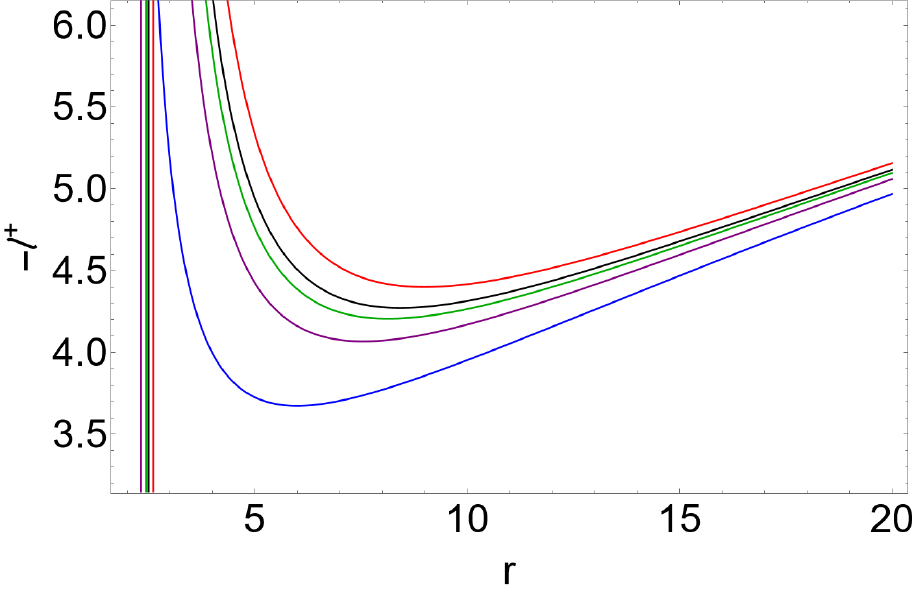}
        \includegraphics[width=8cm]{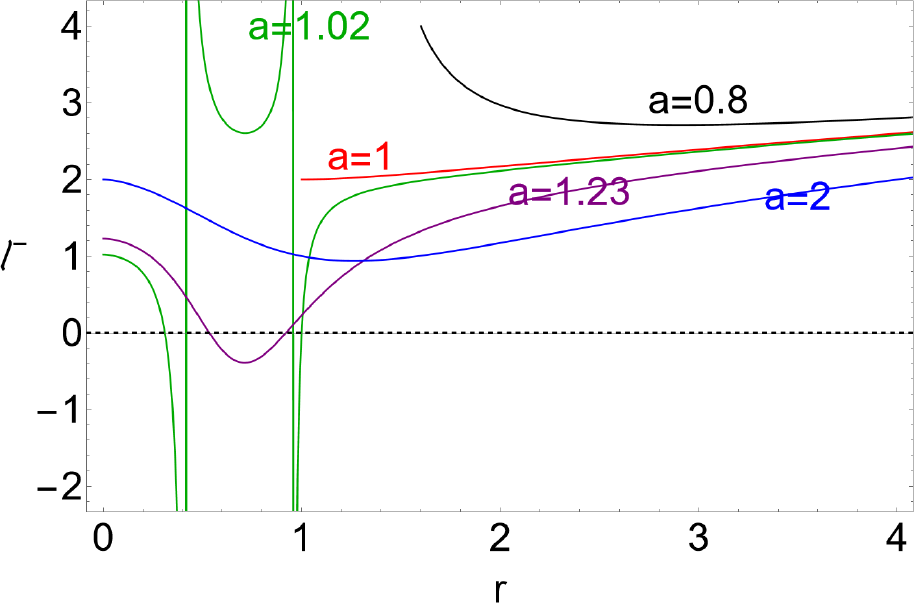}
             \includegraphics[width=8cm]{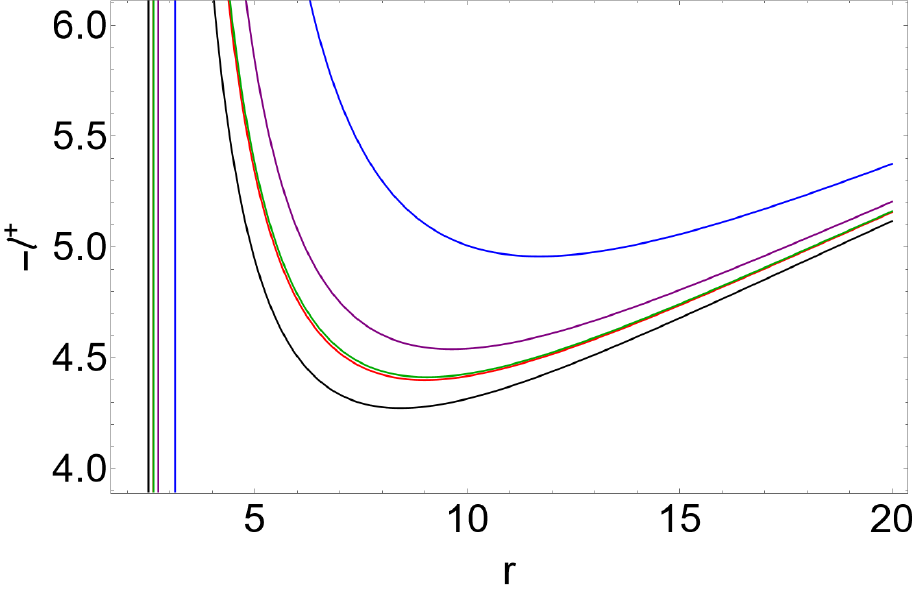}
     \caption{Specific angular momentum $\ell^+$ (right column) and $\ell^-$ (left  column) of Eqs\il(\ref{Eq:pian-l-equa-l}) as functions of radius $r$ for different  Kerr BH spin $a\in[\,0,1\,]$ (upper line) and
     BH and NS spin $a>1$ (bottom line) signed on the curves. Right column panels follow the color code explicit in the right panels. (All quantities are dimensionless).}\label{Fig:PlotdicsprifUaosc}
\end{figure}
{Figs\il(\ref{Fig:PlotdicsprifUaosc}) show fluid specific angular momentum
$\ell^\pm$ as function of $r$ for different spin $a$, for NSs and BHs.}

{
In the {NS} geometries  especially for  small values of the  spin--mass ratio ($a\in]\,1,a_2\,]$ where $a_2\equiv 1.29904$) it is necessary to discuss further the notion of   co-rotating and  counter-rotating motion.}
Considering   quantities  of    $\{ \La, \ell \}$, with $a>0$, we introduce  the following definitions for fluid and test particle motion:
\begin{description}
\item[--]
For a circularly orbiting test particle,    particle counter--rotation  (co-rotation) is \emph{defined} by $\La a<0$ ($\La a>0$).
\item[--]
For  fluids,  the    counter-rotation  (co-rotation) is \emph{defined} by $\ell a<0$ ($\ell a>0$).
\end{description}
We can read  these  definitions\footnote{{The choice to adopt the co-rotation and  counter-rotation classification  for fluids and particles in accordance with the quantities $\ell$ or $\La$ respectively is a  common   choice in literature.  A particle  or fluid orbiting in the NS spacetimes can satisfy $\La a<0$ but also  $\ell=\La/\Em>0$, as there can be $\La a<0$ and $\Em a<0$.
It is therefore important, in the NS spacetimes, to differentiate  the two definitions. (In the Kerr BH spacetime the two definitions are equivalent, as the rotation orientation definition according  $\ell$ or $\La$ are coincident.).
 In  this regard, many aspects of the particles dynamics  are governed  by (and  thus parametrized in accordance with) with the constants  $\La$ and $\Em$, and it is therefore useful in this case to use the  co-rotation and counter-rotation classification according to particle momentum $\La$ for test particles.
 Viceversa,  many aspects of
 fluids  and, more generally, of   the orbiting extended matter,  including those considered in this article, are essentially regulated by (and thus  parametrized in accordance with)  the momentum  $\ell$ (for example, there are accretion disks defined by the parameter $\ell$--see  Sec.(\ref{Sec:extended-geo-struc})). It is therefore natural   to adopt for these systems, differently from  particles,  the rotation orientation  classification according to $\ell$.}}
 in terms of  $\ell$,  $\{\Em,\La\}$ and $\Omega$.
(Note the velocity  $u^\phi$ sign coincides with $\Omega$, assuming   $u^t>0$). )

{For this purpose, it is convenient to introduce radii  $\{r_0^\pm,r_{\delta}^\pm\}$ defined by the conditions that on the equatorial circular  orbits   $r_0^\pm<r_\epsilon^+=2$,   there is   $\La=\ell=0$, while
on  the orbits $r_\delta^\pm: r_{0}^-< r_\delta^-<r_\delta^+< r_0^+$, there is $\Em=0$ and $\La<0$ see Figs\il(\ref{Fig:Plotdicsprosc}) and \cite{Pu:Kerr,2005MPLA...20..561S,slany,1981BAICz..32...68S}.
That is  radii $r_0^\pm$  can be found from the condition $\La=\ell=0$ on the geodesic motion and  $r_\delta^\pm$  can be found from the condition $\Em=0$ on the geodesic motion. Their  explicit form is in  Appendix\il(\ref{Appendix:radii}). From the analysis of  $r_\delta^\pm$ and $r_0^\pm$ we introduce the spins,
\bea&&\label{Eq:spins-a0-a1-a2}
a_0: r_\delta^+=r_\delta^-\equiv \frac{4}{3} \sqrt{\frac{2}{3}}=1.08866,\quad a_2: r_0^+=r_0^-\equiv \frac{3 \sqrt{3}}{4}=1.29904.
\eea
--see Figs\il(\ref{Fig:Plotdicsprosc}). Hence spins $a_0$ and $a_2$ can be found assuming  conditions $r_\delta^+=r_\delta^-$ and  $r_0^+=r_0^-$ respectively, and represent  a limiting  case, where in these NS spacetimes there is only one limiting orbit where $\Em=0$ and  $\La=0$ respectively (Figs\il(\ref{Fig:Plotdicsprosc})). As detailed below $a_0$ and $a_2$  are maximum spin where radii $r_\delta^\pm$ and $r_0^\pm$ exist. }
 Therefore, we can summarize the situation for counter-rotating  flows and co-rotating flows   as follows:
\begin{description}
\item{\textbf{Counter-rotating tori  $(\ell<0)$}}

This case  includes  fluids  with $\ell=\ell^+<0$ and  $\ell=\ell^-<0$ in the ergoregion for {NSs} with $a\in\, [\,1,a_2\,]$,  more precisely   there are:
\begin{itemize}
\item[\textbf{(I)}] Tori with $\ell=\ell^+<0$,  that is with $(\La<0,\Em>0)$ (located far from the ergoregion)
\\
\item[\textbf{(II)}] {Tori with momentum $\ell=\ell^-<0$, that is with $(\La<0,\Em>0)$, in
$]\,r_0^-,r_\delta^-\,[\, \cup\, ]\,r_\delta^+,r_0^+\,[$, for spacetimes $a\in\,[\,1, a_0\,]$, and in the orbital region $]\,r_0^-,r_0^+\,[$, for {NSs} with spin $a\in\,]\,a_0,a_2\,]$--Figs\il(\ref{Fig:Plotdicsprosc}).}
\end{itemize}
\item{\textbf{Co-rotating tori  ($\ell>0$)}}

Co-rotating tori, i.e. with  $\ell=\ell^->0$, are defined for $(r>0,a>a_2)$,   $(r\in\, ]\,0,r_0^-\,[\, \cup\, ]\,r_0^+,+\infty\,[, a\in\, ]\,a_0,a_2\,[\, )$ and  $(r\in\, ]\,0,r_0^-\,[\, \cup\, ]\,r_\delta^-,r_\delta^+\,[\, \cup\,  r>r_0^+,a\in\, ]\,1, a_0\,[\, )$--see Figs\il(\ref{Fig:Plotdicsprosc}).
\end{description}
 \begin{table}
 \resizebox{.991\textwidth}{!}{%
\begin{tabular}{l|l}
  \hline
  $\La$ test particles angular momentum (constant of motion)&
  Eq.(\ref{Eq:EmLdef})
  \\
   $\Em$ test particles energy  (constant of motion)&
  Eq.(\ref{Eq:EmLdef})
  \\
  $\ell\equiv \La/\Em$ specific angular momentum (constant of motion)& Eq.\il(\ref{Eq:flo-adding})
  \\
 $\ell^+$ counter-rotating specific angular momentum &  Eq.\il(\ref{Eq:pian-l-equa-l})--Sec.\il(\ref{Sec:coro-contro})
 \\
  $\ell^-$ co-rotating or counter-rotating specific angular momentum & Eq.\il(\ref{Eq:pian-l-equa-l})--Sec.\il(\ref{Sec:coro-contro})
 \\
$\sigma_\epsilon: r_\epsilon^+=r_\epsilon^-$ &Eq.\il(\ref{Eq:fioc-rumor})
\\
 $r_\delta^\pm: \Em(r_\delta^\pm)=0, \La(r_\delta^\pm)<0$&
 Figs\il(\ref{Fig:Plotdicsprosc})-- Appendix\il(\ref{Appendix:radii})
 \\
 $r_0^\pm: \La(r_0^\pm)=\ell(r_0^\pm)=0$&Figs\il(\ref{Fig:Plotdicsprosc})--Sec.\il(\ref{Sec:coro-contro})--Appendix\il(\ref{Appendix:radii})
\\
$
a_0: r_\delta^+=r_\delta^-\equiv \frac{4}{3} \sqrt{\frac{2}{3}}=1.08866$&
Eqs\il(\ref{Eq:spins-a0-a1-a2})-- Figs\il(\ref{Fig:Plotdicsprosc})
\\
$a_2: r_0^+=r_0^-\equiv \frac{3 \sqrt{3}}{4}=1.29904$&
Eqs\il(\ref{Eq:spins-a0-a1-a2})-- Figs\il(\ref{Fig:Plotdicsprosc})
\\
$
{r}_{[mb]}^{\pm}:\;\ell^{\pm}(r_{mb}^{\pm})=
 \ell^{\pm}({r}_{[mb]}^{\pm})\equiv {\ell_{mb}^{\pm}}$ & Eq.\il(\ref{Eq:conveng-defini})--Figs\il(\ref{Fig:Plotdicsprosc})
 \\
$
  r_{[\gamma]}^{\pm}: \ell^{\pm}(r_{\gamma}^{\pm})=
  \ell^{+}(r_{[\gamma]}^{+})\equiv \ell_{\gamma}^{+}$& Eq.\il(\ref{Eq:conveng-defini})--Figs\il(\ref{Fig:Plotdicsprosc})
   \\
$a_1 \equiv 1.28112:
\bar {r} _ {ms}^ -= \tilde {r} _ {ms}^-$ & Sec.\il(\ref{Sec:extended-geo-struc})
   \\
$r_\Ta^\mp(a,\sigma;\ell): u^\phi=0$ inversion radius& Eq.\il(\ref{Eq:rinversionmp})--
Figs.\il(\ref{Fig:Plotvinthallo})--
Figs.\il(\ref{Fig:Plotvinthalloa})-- Figs.\il(\ref{Fig:Plotsigmalimit3W})
 \\
 $\sigma_\Ta(a,r;\ell): u^\phi=0$ inversion plane& Eq.\il(\ref{Eq:rinversionmp})
 \\
 $()_\Ta$ any quantitiy $()$ evaluated on the inversion surface & Sec.\il(\ref{Sec:flow-inversion-points})
 \\
 $r_{cr}: r_\Ta^-= r_\Ta^+$ & Eq.\il(\ref{Eq:rcorss-defs})--
--Figs\il(\ref{Fig:Plotvinthallosrc12W}),(\ref{Fig:plotrcross}),(\ref{Fig:Plotvinthallo}),(\ref{Fig:Plotvinthalloa})
\\
$ r_{\Ta\max}^0: \partial_a \sigma_\Ta=0$ & Eq.\il(\ref{Eq:rmax})--Figs\il(\ref{Fig:Plotsigmamax})
 \\
$r_{\Ta\max}^\pm: \partial_{r_\Ta}\sigma_\Ta=0$&Eq.\il(\ref{Eq:rmax})--Figs\il(\ref{Fig:Plotsigmamax})
\\
$\ell_e^-:\partial_a r_\Ta^\pm=0$& Eqs.\il(\ref{Eq:merka.til-vant-sal-qua-maxmin}), Eqs.\il(\ref{Eq:lemp})--Figs\il(\ref{Fig:Plotsigmalimit3}).
\\
\hline
  \end{tabular}}
  \caption{{Lookup table containing the main notation and quantities used throughout this paper.  Links to associated sections, definitions or  figures are also listed.}
}\label{Table:Notation}
  \end{table}
  {
There is therefore   $\ell^-<0$ with  $(\La<0,\Em>0)$  in the {NS} ergoregion,   together with   co-rotating solutions  $\ell^->0$  with $(\La<0,\Em<0)$ with negative energy $\Em$.  However, these solutions  correspond to  the relativistic angular velocity, i.e. the Keplerian velocity with respect to static observers at infinity,    $\Omega>0$. In this sense, they are all co-rotating with respect to the static observers at infinity.}
\subsection{Extended geodesic structure and tori constraints}\label{Sec:extended-geo-struc}
{Test particle motion and many aspects of the physics of accretion around compact objects is constrained by the  spacetime geodesic structure. The flow dynamics  we consider in this analysis is regulated by the set of co-rotating and counter-rotating Kerr NSs circular  equatorial geodesics  radii. This set of radii defines  and constrains different extended orbiting surfaces, as accretion tori having  different topology and  their instability points. It is therefore useful to consider here in details the spacetime co--rotating and counter--rotating  geodesic structure. }

The  Kerr {NS} background    geodesic structure is constituted by the
radii $\{r_{\gamma}^{\pm},r_{mb}^{\pm},r_{ms}^{\pm}\}$,  marginally circular (photon) orbit, marginally bounded orbit,  and marginally stable  orbit respectively,  regulating the location of  the   tori and  proto-jets  cusps.
{(For convenience we report in Appedix\il(\ref{Appendix:radii})  the explicit expression of  the radii of the Kerr NS spacetime geodesc structure).}
The
  geodesic  structure is extended  to include      radii  $\{r_{[\gamma]}^{\pm},r_{[mb]}^{\pm},r_{[ms]}^{\pm}\}$  regulating  the location of the   toroidal configurations centers
:
\bea&&\label{Eq:conveng-defini}
{r}_{[mb]}^{\pm}:\;\ell^{\pm}(r_{mb}^{\pm})=
 \ell^{\pm}({r}_{[mb]}^{\pm})\equiv {\ell_{mb}^{\pm}},
\quad
  r_{[\gamma]}^{\pm}: \ell^{\pm}(r_{\gamma}^{\pm})=
  \ell^{+}(r_{[\gamma]}^{+})\equiv \ell_{\gamma}^{+},
 \\ &&\label{Eq:pre-Nob}
\mbox{where}\quad r_{\gamma}^{\pm}<r_{mb}^{\pm}<r_{ms}^{\pm}<
 {r}_{[mb]}^{\pm}\quad\mbox{and}\quad
 {r}_{[mb]}^{+}< r_{[\gamma]}^{+}<r_\gamma^-\equiv 0,
\eea
respectively where there is   $r_\gamma^-\equiv 0$.
%.
 Increasing the {NS} spin, radii of geodesic structure and  tori  depart from the central singularity,    for both co-rotating and counter-rotating fluids, constituting a major difference with the  {BH}  case.
Note there is  $r_{[ms]}^+=r_{ms}^+$ where $r_{[ms]}^+: \ell_{ms}^+\equiv\ell(r_{ms}^+)=\ell^+(r)$, analogously a similar relation holds for fluids with $\ell^-$  at {NSs} spins $a>a_2$.
\begin{figure}
   \includegraphics[width=5.6cm]{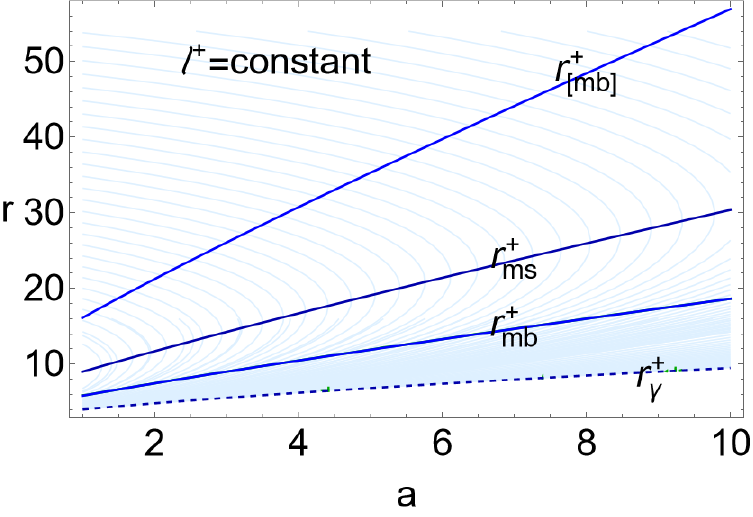}
    \includegraphics[width=5.6cm]{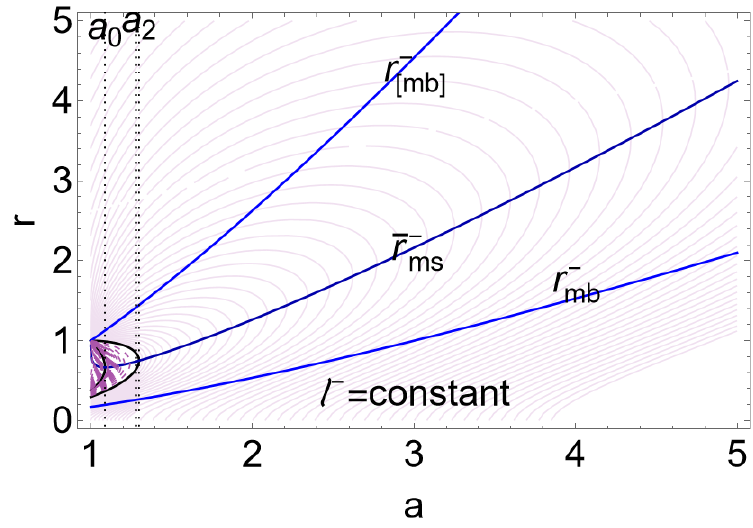}
        \includegraphics[width=5.6cm]{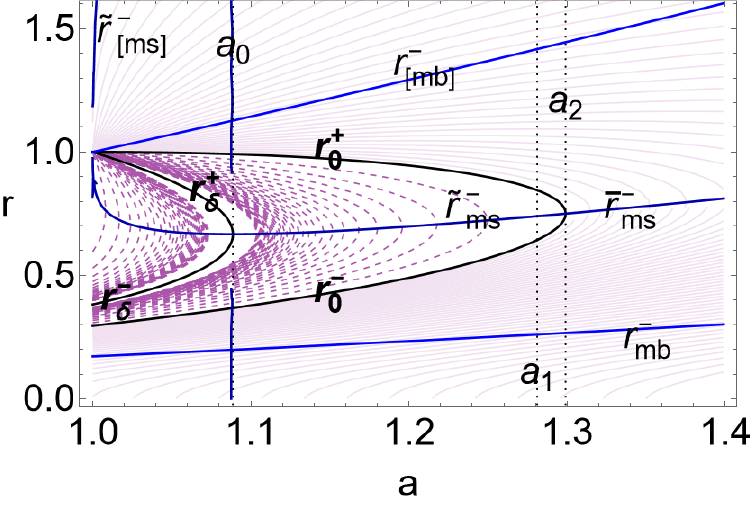}
     \caption{Analysis of co-rotating and counter-rotating tori. Curves $\ell^-=$constant (right  and center panel) and $\ell^+$=constant (left panel)  ($\ell^+<0$ and $\ell^-\lesseqgtr0$  are the fluid specific angular momentum) in the plane $r-a$. (All quantities are dimensionless).
    Right  panel is a zoom of the center panel.  $r_\gamma^\pm$ are the last circular orbit  (there is $r_\gamma=0$).
    There is  $r_{0}^\pm:\La=\ell=0$.
     Radii $r_{mb}^\pm$  are the marginally bounded orbits. Spins $\{a_0,a_2\}$ are defined in Eqs\il(\ref{Eq:spins-a0-a1-a2}), $a_1 \equiv 1.28112:
\bar {r} _ {ms}^ -= \tilde {r} _ {ms}^-$ is defined in Sec.\il(\ref{Sec:extended-geo-struc}). Radii $r_{ms}^-=\{\tilde{r}_{ms}^-,\bar{r}_{ms}^-\}$ and $r_{ms}^+$ are the marginally stable orbits for the fluid with specific angular momentum $\ell^-$ and $\ell^+$ respectively defined in Eqs\il(\ref{Eq:pian-l-equa-l}). Radii $r_{[mb]}^{\pm}$ are  in Eqs\il(\ref{Eq:conveng-defini}).  {Radii of the geodesic structures are  in Appendix\il(\ref{Appendix:radii})}.}\label{Fig:Plotdicsprosc}
\end{figure}
The situation for fluid with momentum $\ell^-$ is however  more complex.
There is $\ell^-\lesseqgtr0$, constrained  by the radii $\{r_{mb}^-,\tilde{r}_{ms}^-,\bar{r}_{ms}^-,r_\gamma^-\}$, with  $\{\La\lesseqgtr0,\Em\lesseqgtr0\}$.
Radii $r_{ms}^-=\{\bar{r}_{ms}^-,\tilde{r}_{ms}^-\}$, are the marginally stable orbits related to the motion  $\ell^-\gtrless 0$, where $
r_{ms}^-=\bar {r} _ {ms}^- $ for  $a>a_1$ and $r_{ms}^-=
\tilde {r} _ {ms}^-$  for  $a\in\,]\,1, a_1\,]$  and {$
\bar {r} _ {ms}^ -= \tilde {r} _ {ms}^-$ on  $a_1 \equiv 1.28112$--Figs\il(\ref{Fig:Plotdicsprosc})}.
However, for simplicity of notation, when it is not necessary to specify,
we will use the abbreviated notation  $r_{ms}^-$ for $\{\bar{r}_{ms}^-,\tilde{r}_{ms}^-\}$.

Barotropic counter-rotating closed quiescent (not cusped tori, regular topology), cusped or open configurations have centers coincident with maximum of pressure and density point in the disk, the  tori cusp is the minimum point of pressure and density.

Tori with  specific angular momentum $\ell=\ell^+<0$ are constrained as follows \cite{Pu:Kerr}
\begin{description}
\item[--]
For $\ell^+\in \, \, \mathbf{L_1^+}\equiv\,]\,\ell^+_{mb}, \ell^+_{ms}\,[$ there are quiescent  and cusped tori. The cusp  is  $r_{\times}\in \,]\,r_{mb}^+,r_{ms}^+\,]$, and  the torus center with maximum pressure is  $r_{center}\in \,]\,r_{ms}^+,r_{[mb]}^+\,]$;
\item[--]
%\hline
For $\ell^+\in \, \mathbf{L_2^+}\equiv\,[\, \ell^+_{\gamma},\ell^+_{\mb}\,[$ there are  quiescent  tori and proto-jets.  Proto-jets are  open cusped equipotential  surfaces. Unstable point, cusp, is on  $r_{j}\in \,]\,r_{\gamma}^+,r_{mb}^+\,]\,$  and  center with maximum pressure $r_{center}\in \,]\,r_{[mb]}^+,r_{[\gamma]}^+\,]\,$;
\item[--]
%\hline
For $\ell^+\in \, \mathbf{L_3^+}: \ell^+<\ell^+_{\gamma}$, there are quiescent  tori,  with center $r_{center}>r_{[\gamma]}^+$.
\end{description}
A very large $\ell$ in magnitude is  typical of proto-jets  or  quiescent   toroids orbiting  far from the central attractor for counter-rotating tori (with $\ell=\ell^+<0$)   or some co-rotating tori with $\ell=\ell^->0$) in a class of slow rotating {NSs}.

A more articulated structure characterizes {NSs} with  spin $a\in\, ]\,1,a_2\,]$ and fluids with $\ell^-$  on the equatorial plane ergoregion.
For $a>a_2$, where $\Em>0$ and $\La>0$  $(\ell^->0)$, the  geodetical orbital structure for $\ell^-$ is similar to the counter-rotating tori with $\ell=\ell^+$, where $r_\gamma^-=0$. {For $a\in \, [\,1,a_0\,[$, the $\ell^-$ geodesic structure is more articulated.
For  $a\in\, [\,1, a_2\,]$, there can be    tori   with $\ell^-<0$ and ($\Em>0,\La<0$) in the ergoregion  and
  tori with  $\ell^->0$ and ($\Em<0,\La<0$). There can be  double tori system  with $\ell=\ell^+=\ell^-<0$ or $\ell=\ell^->0$. For tori with  centers and  cusps located on  $r_{0}^\pm$   there is $\ell=\La=0$,  and  tori in the  limiting case of $\ell=0$ are considered   in Sec.\il(\ref{Sec:inversionlequal0}). On  $r_\delta^{\pm}$ ("center" and "cusp" respectively)  there is  $\Em=0$ ($\ell$ is not well defined).}
\section{Flow inversion points}\label{Sec:flow-inversion-points}
{In this section the inversion points and inversion surfaces definitions are introduced and we  start  with the analysis of the inversion  surfaces  extremes and limits focusing, in particular, on the location of the inversion points with the respect to the central singularity.}

 The flow inversion points  are defined by the condition   $u^{\phi}=0$ on the flow particle velocities, i.e. as points of vanishing  axial velocity of the flow motion as related to distant static observers.

From the definition of constant $\ell$  and  $\Em$, Eqs\il(\ref{Eq:EmLdef}) and  Eq.\il(\ref{Eq:flo-adding}),  fixed by the orbiting torus, and the  inversion point definition   we obtain:
\bea\label{Eq:lLE}
&&\ell=-\left.\frac{g_{t\phi}}{g_{tt}}\right|_\Ta=-\frac{2 a r_{\Ta} \sigma_{\Ta} }{\Sigma_\Ta -2 r_\Ta},
\quad \mbox{where}
\quad
\Em_\Ta \equiv - g_ {tt}(\Ta) \dot {t}_\Ta,\quad \La_\Ta \equiv g_ {t\phi}(\Ta)\dot{t}_\Ta,
\eea
where $\Em_\Ta$ and  $\La_\Ta$ are the energy   and momentum of the flow particles at the inversion point. 
 More in general we adopt the notation $q_{\bullet}\equiv q(r_{\bullet})$ for any quantity $q$ evaluated on a radius $r_{\bullet}$.  Therefore,    notation $q_\Ta$ or $q(\Ta)$ is for  any quantity $q$ considered at the inversion point, and $q_0=q(0)$ for  any  quantity $q$ evaluated at the initial point of the (free-falling) flow trajectories.

These quantities are independent from  the normalization condition,  being a consequence of the definition of constant  $\{\ell,\mathcal{E},\laa\}$ only.

Using Eq.\il(\ref{Eq:lLE}), we can discuss   the existence of the inversion points radii $r_\Ta^\pm$ and planes $\sigma_\Ta$, for co-rotating   $\ell>0$ or counter-rotating $\ell<0$ fluids,   with constant $\ell$, introducing the quantities
\bea\label{Eq:rinversionmp}
&&r_\Ta^\mp=\mp\sqrt{a^2 \left(\sigma_\Ta +\frac{\sigma_\Ta ^2}{\ell ^2}-1\right)-\frac{2 a \sigma_\Ta }{\ell }+1}-\frac{a \sigma_\Ta }{\ell }+1,\quad\mbox{and}
\quad
\sigma_\Ta\equiv\frac{\ell \Delta_\Ta}{a (a \ell -2 r_\Ta)}.
\eea
where $
\Delta_\Ta$ is $\Delta$  defined at the inversion point--see Fig.\il(\ref{Fig:Plotsigmalimit3W}).
\begin{figure}
\centering
        \includegraphics[width=8cm]{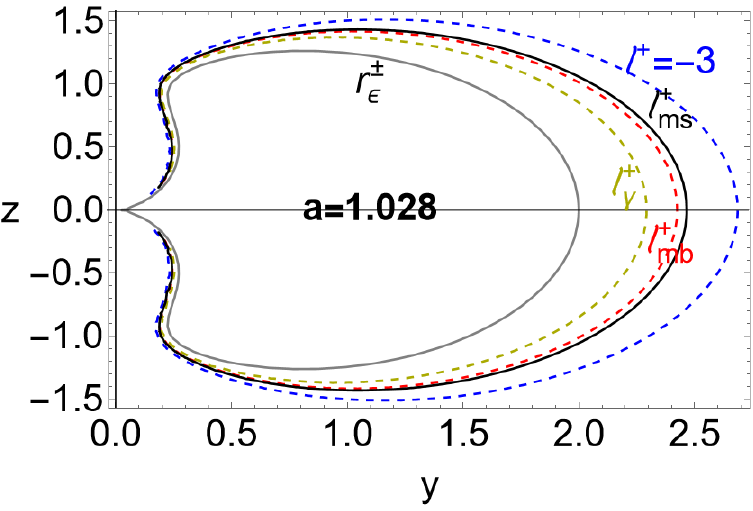}
       \caption{Inversion surfaces $r_\Ta^{\pm}(\ell)$ in the Kerr  NS  spacetime with   spin-mass ratio $a=1.028$, at different  counter-rotating specific angular momentum  $\ell$  for $\ell\in\{\ell_\gamma^+,\ell_{mb}^+,\ell_{ms}^+,-3\}$ as signed on the curves.   For $a=1.028$ there is
       $\ell_{\gamma}^+=-7.04659$ (yellow curve),  $\ell_{mb}^+=-4.84816$ (red curve), $\ell_{ms}^+=-4.4173$ (black curve).  Hence $\ell_{\gamma}^+< \ell_{mb}^+<\ell_{ms}^+<-3$ and  $r_\epsilon^\pm<r_\Ta(\ell_{\gamma}^+)< r_\Ta(\ell_{mb}^+)<r_\Ta(\ell_{ms}^+)<r_\Ta(-3)$.  Gray  curves  are for radii  $r_{\epsilon}^+>r_{\epsilon}^-$, i.e.  the outer and inner ergosurfaces respectively.
       All the quantities are dimensionless. Notation $ms$ refers the marginally stable circular orbit, $mb$ to marginally bounded orbit and $(\gamma)$ to the marginally circular orbit.  There is $r=\sqrt{z^2+y^2}$ and $\sigma=y^2/(z^2+y^2)$, for  $\sigma\equiv\sin^2\theta$ (where $\sigma=1$ is the equatorial plane).
  }\label{Fig:Plotsigmalimit3W}
\end{figure}
There is
\bea&&\label{Eq:rcorss-defs}
 r_\Ta^-= r_\Ta^+=r_{cr}\equiv a\sqrt{1-\sigma_\Ta}\quad\mbox{on}\quad \ell_{cr}=\frac{a\sigma_\Ta[r_{cr}+1]}{1-r_{cr}^2}.
\eea
There are no timelike or photon-like inversion points for $\ell>0$,  as clear from the study of the conditions $(\Em,\La)$ constant and normalization condition.
Radius $r_{cr}$ is shown in  Figs\il(\ref{Fig:Plotvinthallosrc12W}),(\ref{Fig:plotrcross}),(\ref{Fig:Plotvinthallo}),(\ref{Fig:Plotvinthalloa}).
\begin{figure}
                       \includegraphics[width=5.75cm]{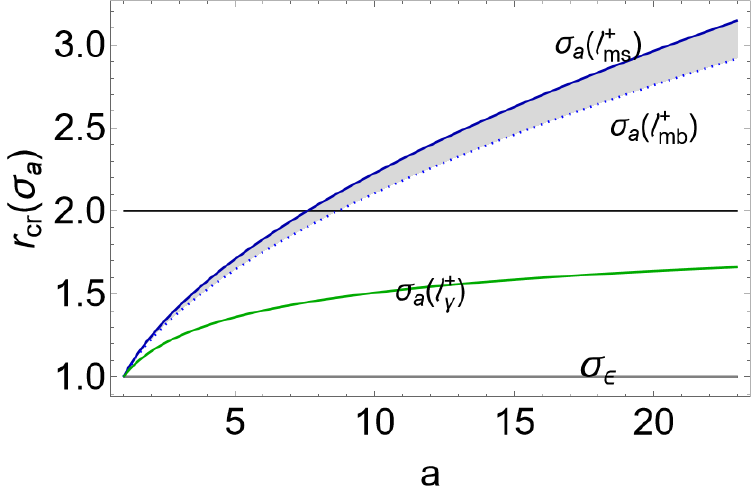}
   \includegraphics[width=5.75cm]{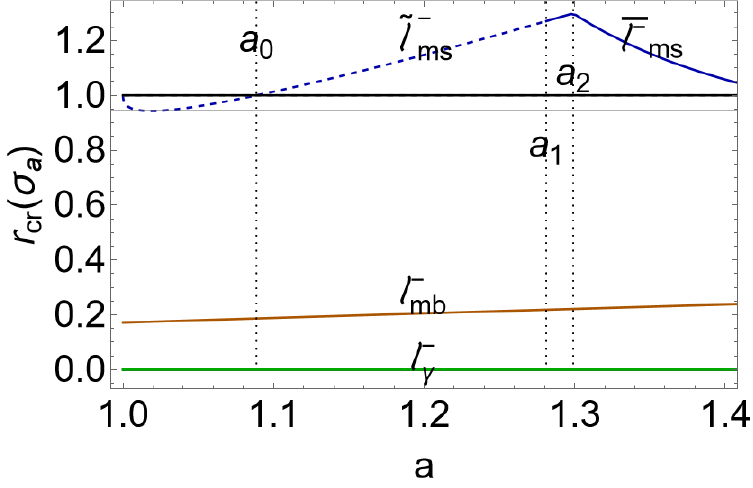} \includegraphics[width=5.75cm]{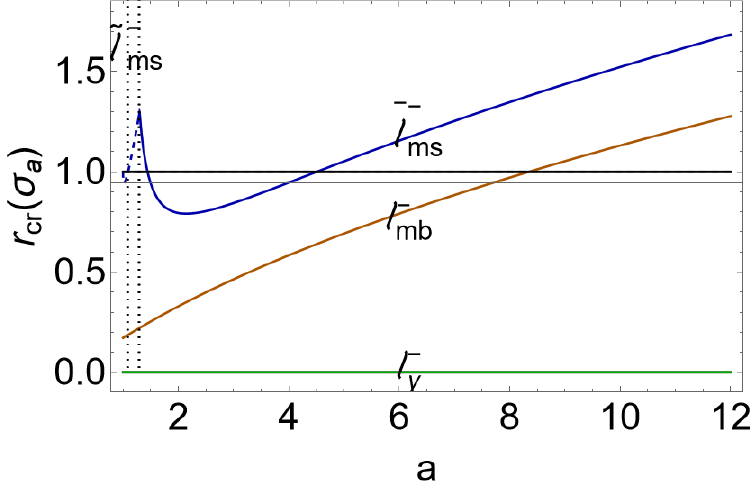}
     \caption{ Panels show radius $r_{cr}: r_{\Ta}^-=r_\Ta^+$ (flow  inversion point radius) defined in Eq.\il(\ref{Eq:rcorss-defs})  on  the plane $\sigma_{a}(\ell)$ for different fluid specific angular momenta $(\ell)$, as function of the {NS}  spin $a>1$.   Left panel shows the  analysis for  the counter-rotating fluid specific angular momentum $\ell^+<0$.  Center and  right panels show   the analysis for the momentum $\ell^-$.
  Central panel is a close-up view of the right panel.
      Plane $\sigma_\epsilon$ is defined in Eq.\il(\ref{Eq:fioc-rumor}), the plane $\sigma_{a}$ is defined in Eq.\il(\ref{Eq:sasW}). The ergoregion is defined for $\sigma\in[\,\sigma_\epsilon,1\,]$. There is $\sigma\equiv \sin^2\theta\in[\,0,1\,]$, where $\sigma=1$ is the equatorial plane.
      Notation $(mb)$ is for quantities evaluated on the marginally bounded orbit, $(ms)$ for marginally stable orbit, $(\gamma)$ for marginally circular orbit.  Radius $r=r_{\epsilon}^+=2$ is the outer-ergosurface on the equatorial plane.  Spins $\{a_0,a_2\}$, vertical dotted lines, are defined in Eqs\il(\ref{Eq:spins-a0-a1-a2}) and $a_1 \equiv 1.28112:
\bar {r} _ {ms}^ -= \tilde {r} _ {ms}^-$ is defined in Sec.\il(\ref{Sec:extended-geo-struc}).  There is $r_{ms}^{-}\equiv\{\tilde{r}_{ms}^-,\bar{r}_{ms}^-\}$, respectively there is $\{\tilde{\ell}_{ms}^-,\bar{\ell}_{ms}^-\}$--see  Sec.\il(\ref{Sec:extended-geo-struc}). Here the most general solutions $r_\Ta^\pm$ of  Eqs\il(\ref{Eq:rinversionmp}) are shown. (All quantities are dimensionless).} \label{Fig:Plotvinthallosrc12W}
\end{figure}
\begin{figure}
  \includegraphics[width=5.75cm]{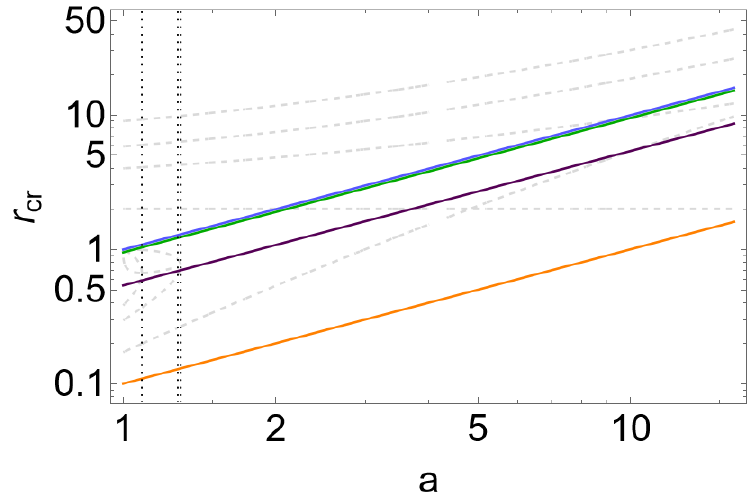}
   \includegraphics[width=5.75cm]{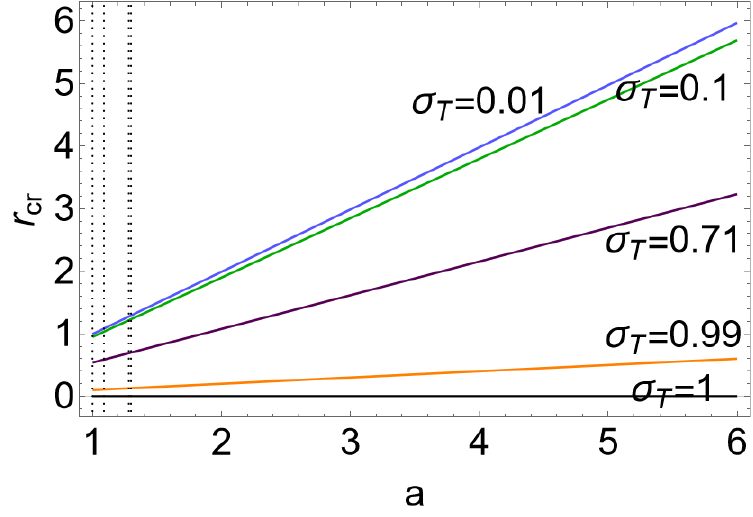}
   \includegraphics[width=5.75cm]{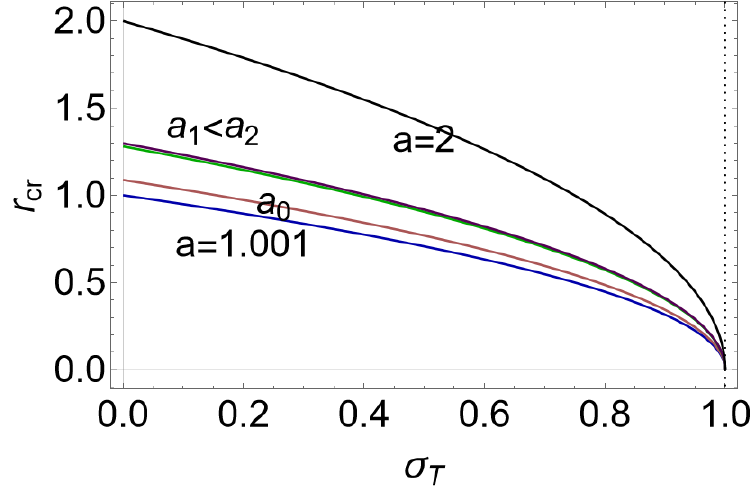}
 \caption{Analysis of counter-rotating flow inversion points $r_\Ta^{\pm}$ with fluid specific angular momentum $\ell^+<0$. (Most general solutions $r_\Ta^\pm$ are shown, discussion on further constraints on the constants of motion is in Sec.\il(\ref{Sec:all-to-rediid}).)  Left panel shows in light-gray dashed curves the geodesic structure of the {NSs} as in Figs\il(\ref{Fig:Plotdicsprosc}). Dotted vertical lines are  the spins $a_0<a_1<a_2$.  Spins $\{a_0,a_2\}$ are defined in Eqs\il(\ref{Eq:spins-a0-a1-a2}) and  $a_1 \equiv 1.28112:
\bar {r} _ {ms}^ -= \tilde {r} _ {ms}^-$ is defined in Sec.\il(\ref{Sec:extended-geo-struc}). Radius $r_{cr}=r_\Ta^{\pm}$ of Eq.\il(\ref{Eq:rcorss-defs}) is shown as function of the {NS} spin-mass ratio $a$, at different planes $\sigma\equiv\sin^2\theta$ signed on the  curves (left and center panels), where $\sigma=1$ is the equatorial plane, and as function of  the plane $\sigma\in[\,0,1\,]$, for different spins, signed on the curves (right panel). (All quantities are dimensionless).}\label{Fig:plotrcross}
\end{figure}
 \begin{figure}
\centering
    \includegraphics[width=5.75cm]{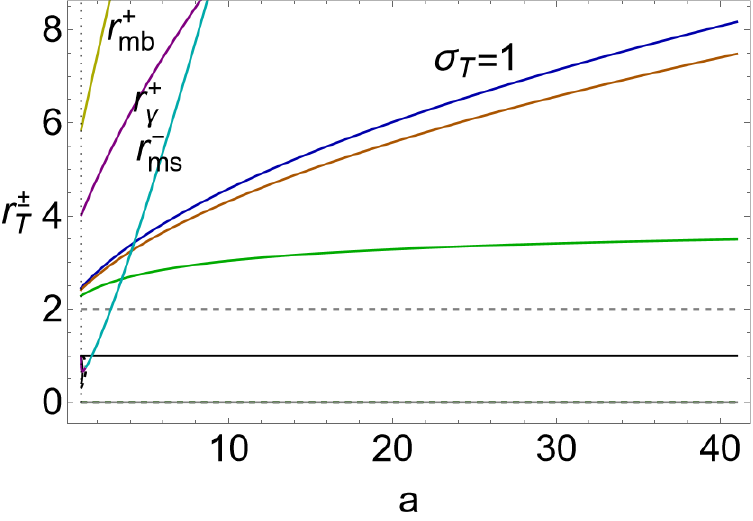}
    \includegraphics[width=5.75cm]{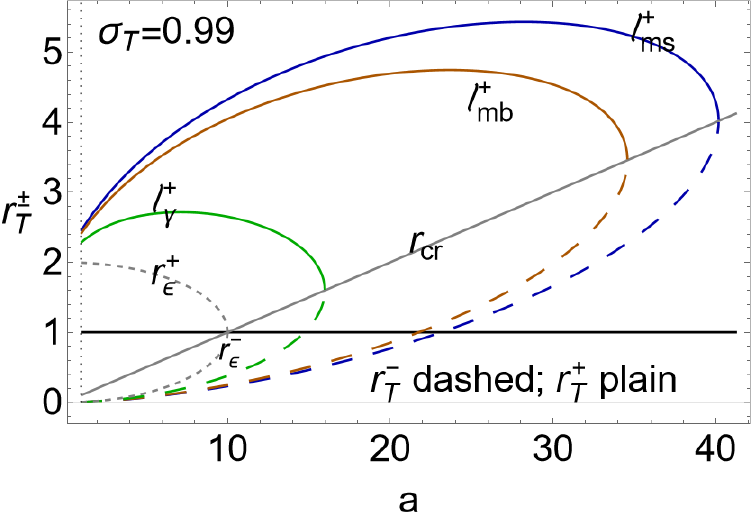}
     \includegraphics[width=5.75cm]{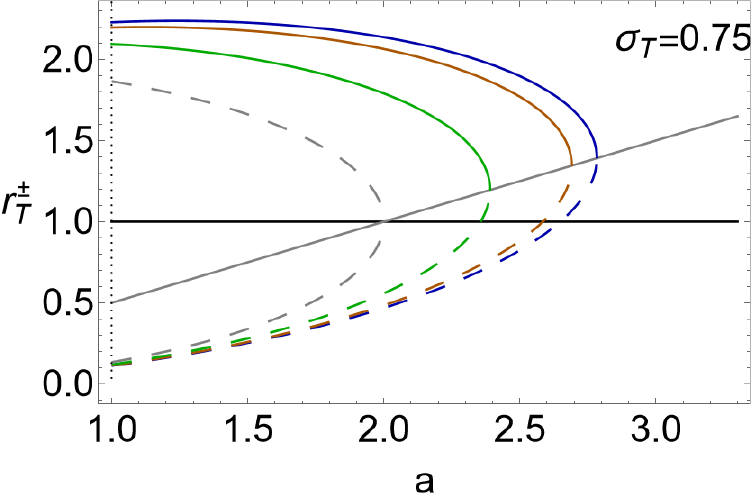}
      \includegraphics[width=5.75cm]{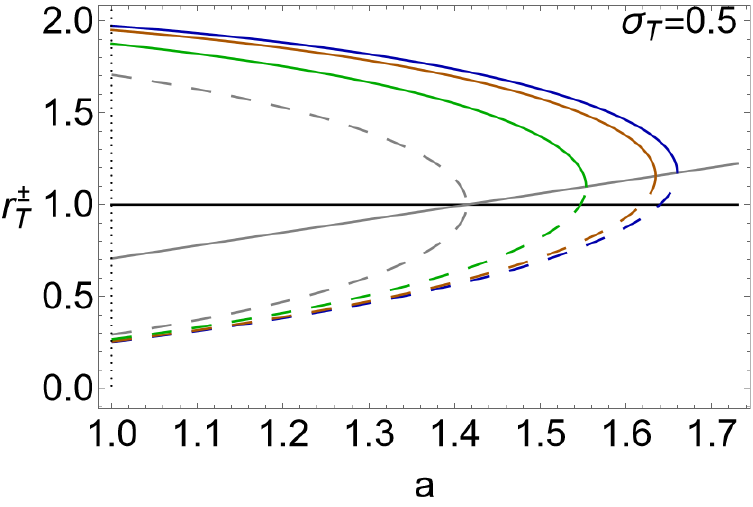}
       \includegraphics[width=5.75cm]{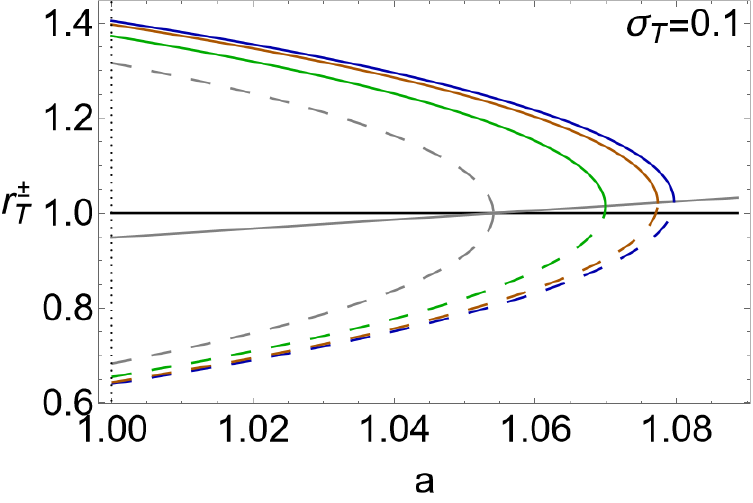}
              \includegraphics[width=5.75cm]{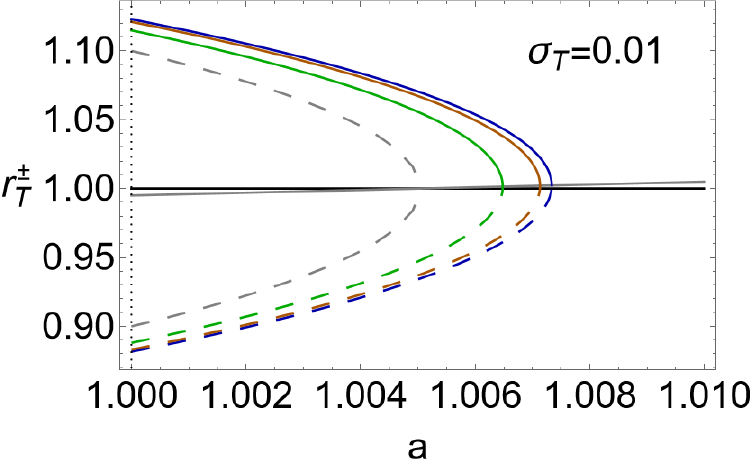}
  \caption{Inversion points $r_\Ta^{\pm}$ of Eqs\il(\ref{Eq:rinversionmp}) are shown  as functions of the {NS} spin-mass ratio $a$, for different planes $\sigma\equiv \sin^2\theta$ signed on the panels, where $\sigma=1$ is the equatorial plane.   (Here the most general solutions $r_\Ta^\pm$ are shown, while discussion on the constants of motion  is in Sec.\il(\ref{Sec:all-to-rediid}).) Dashed gray curves are the outer and the inner ergosurfaces $r_\epsilon^+\geq r_{\epsilon}^-$ respectively.  Plain curves are $r_\Ta^+$, while $r_\Ta^-$ are dashed curves.
  {Curves $r_\Ta^\pm(\ell)$ for $\sigma=1$ are generally non linear functions of the spin $a$ when  evaluated on the function of the spin $\ell=\ell^+(a)\in \,\{\ell_{ms}^+,\ell_{mb}^+,\ell_\gamma^+\}$.}  The radii are evaluated
  for the counter-rotating fluid specific angular momentum $\ell=\ell^+<0$, on the marginally stable orbit $(ms)$-(darker-blue curve), marginally bounded orbit $(mb)$--(blue-curves)  and marginally circular orbit $r_\gamma^+$ (light-blue curves). Radius $r_{cr}=r_\Ta^{\pm}$ (plain gray line)  is defined in  Eq.\il(\ref{Eq:rcorss-defs}). Radius $r_{ms}^-$ is the marginally stable orbit for fluid with $\ell=\ell^-$. (All quantities are dimensionless). }\label{Fig:Plotvinthallo}
\end{figure}
 \begin{figure}
\centering
    \includegraphics[width=5.75cm]{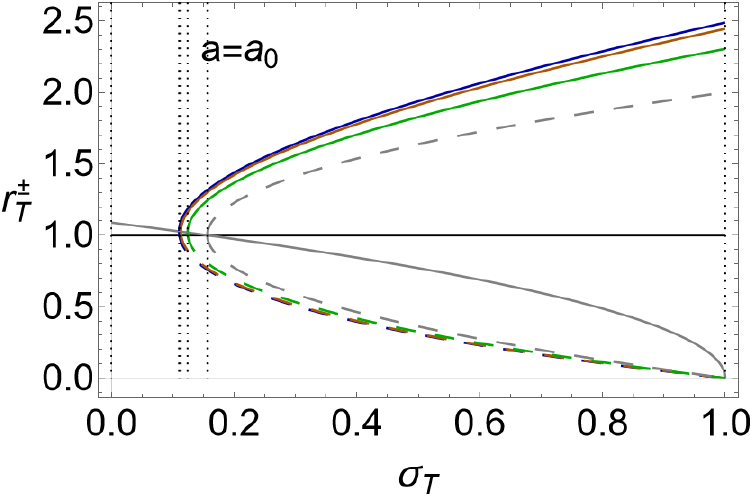}
          \includegraphics[width=5.75cm]{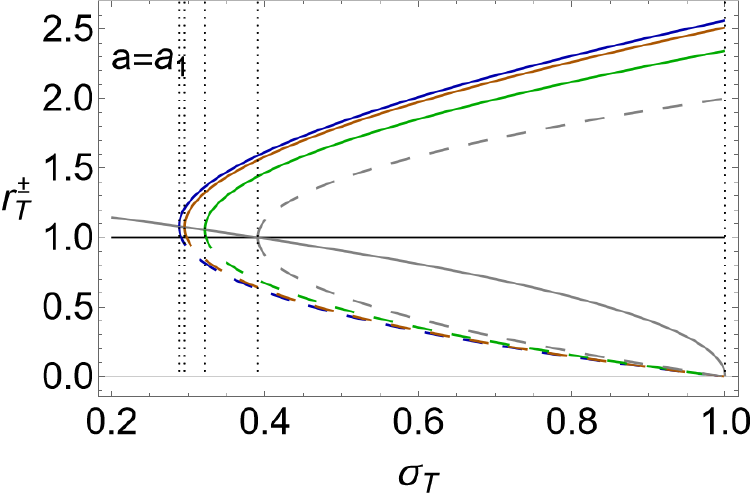}
       \includegraphics[width=5.75cm]{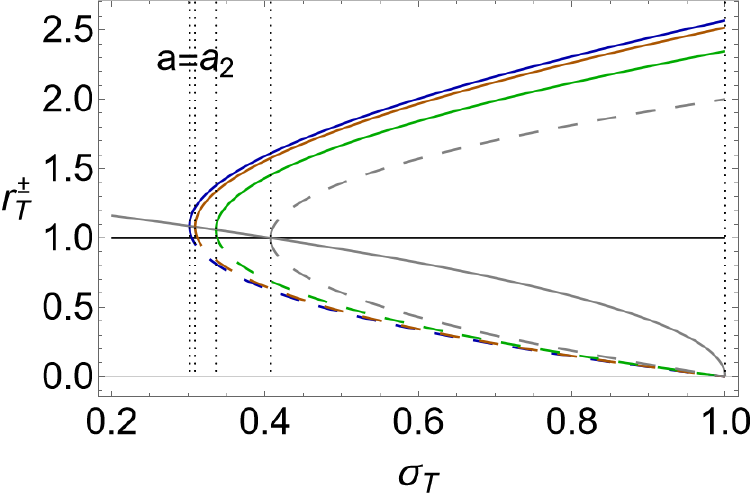}
              \includegraphics[width=5.75cm]{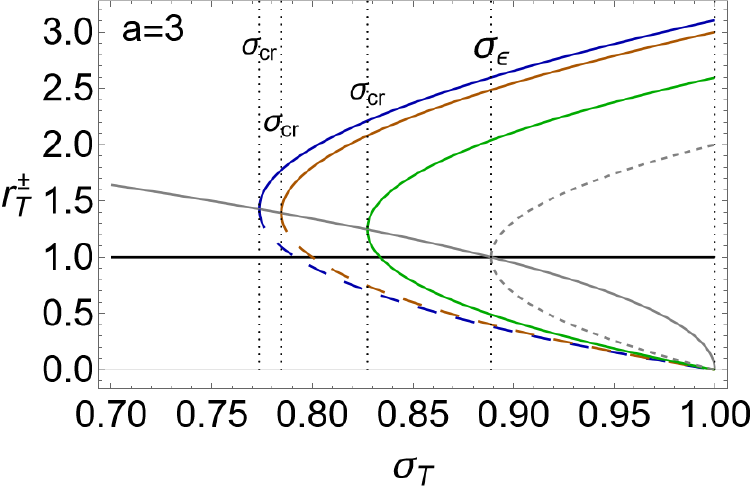}
               \includegraphics[width=5.75cm]{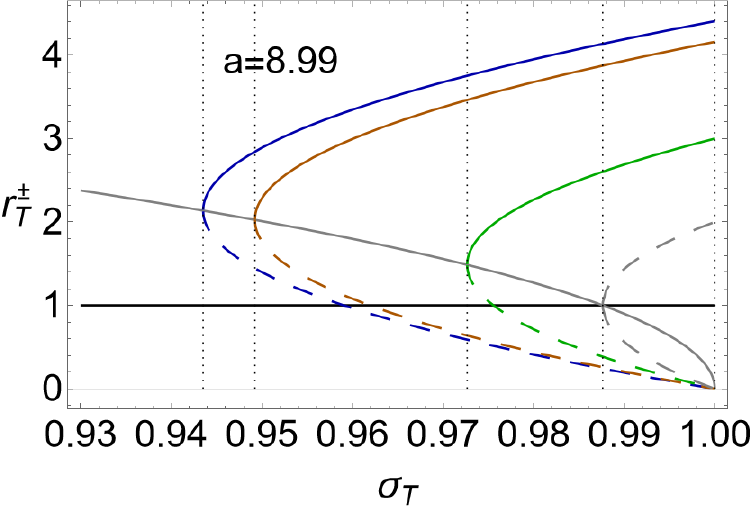}
               \includegraphics[width=5.75cm]{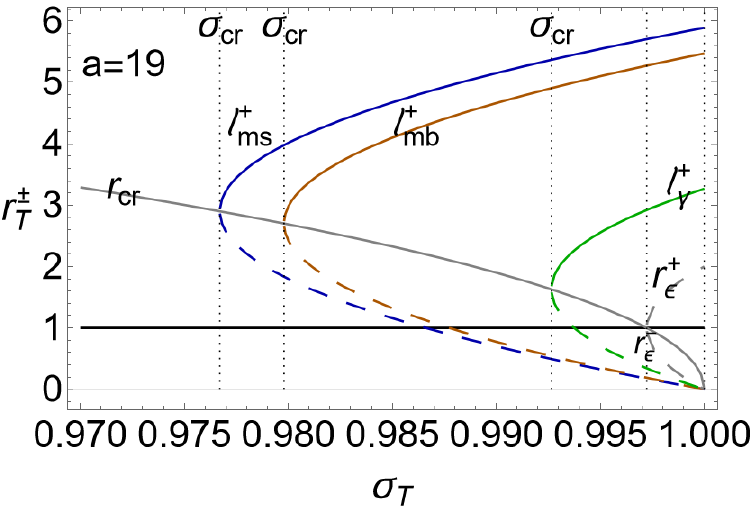}
  \caption{Inversion points $r_\Ta^{\pm}$ of Eqs\il(\ref{Eq:rinversionmp})  of the counter-rotating  flows with specific angular momentum $\ell^+<0$ as functions of the  planes $\sigma\equiv \sin^2\theta$, for different  {NS} spin-mass ratio $a$ signed on the panels, where $\sigma=1$ is the equatorial plane.  (All quantities are dimensionless). Plain curves are $r_\Ta^+$, while $r_\Ta^-$ are dashed curves.
 Spins $\{a_0,a_2\}$ are defined in Eqs\il(\ref{Eq:spins-a0-a1-a2}) and $a_1 \equiv 1.28112:
\bar {r} _ {ms}^ -= \tilde {r} _ {ms}^-$ is defined in Sec.\il(\ref{Sec:extended-geo-struc}).  Limiting plane $\sigma_{cr}$ is function $\sigma_{a}$ of Eqs\il(\ref{Eq:sasW}) evaluated on the selected fluid specific angular momentum. Dashed gray curves are the outer and the inner ergosurfaces $r_\epsilon^+\geq r_{\epsilon}^-$ respectively.  Inversion points are evaluated at specific angular momentum on the marginally stable orbit $(ms)$-(darker-blue curve), marginally bounded orbit $(mb)$--(blue-curves)  and marginally circular orbit $r_\gamma^+$ (light-blue curves).  See also  Figs\il(\ref{Fig:Plotvinthallo}). (The most general solutions $r_\Ta^\pm$ are shown not considering the constants of motion  analysis of Sec.\il(\ref{Sec:all-to-rediid}).)}\label{Fig:Plotvinthalloa}
\end{figure}
Quantities  $(\sigma_\Ta (r_\Ta),r_{\Ta}(\sigma_{\Ta}))$ depend on the  constant of motion  $\ell$ only\footnote{Inversion radius $r_\Ta$ and plane $\sigma_\Ta$ of Eqs\il(\ref{Eq:lLE}) are not independent variables, and they can be found solving the equations of motion or using   further assumptions at any other points of the fluid trajectory.},
describing    both matter and photons,  are independent from the initial particles velocity  ($\{\dot{\sigma}_{\Ta},\dot{r}\}$, therefore their dependence on the tori models and accretion process is limited to the dependence on the fluid specific angular momentum $\ell$, and the results considered here are adaptable to a variety of different general relativistic accretion models.
 Function   $r_\Ta=r_\Ta^\pm(\sigma)$ (equivalently $\sigma_\Ta(r_\Ta)$) defines an almost continuous locus of  inversion points, surrounding the central singularity, \emph{inversion surface}, where   the condition  $u^{\phi}=0$ is satisfied.  For flows from the orbiting tori  there is a maximum and a minimum boundary  $r_\Ta(a;\ell,\sigma_\Ta)$, associated to a maximum and minimum value of $\ell$. This region, as well as  its boundaries, will  be   called the \emph{inversion corona}.
 The extremes of $\ell$ parameters are determined by the tori  parametrized with $\ell$.
 Therefore,  the accretion driven  inversion corona has boundaries defined by $r_\Ta^\pm$ (or $\sigma_\Ta$), evaluated on $\ell_{ms}^\pm$ and $\ell_{mb}^\pm$, while the proto-jets driven coronas has (in general) boundaries defined by $r_\Ta^\pm$ (or $\sigma_\Ta$), evaluated on $\ell_{mb}^\pm$ and $\ell_{\gamma}^\pm$--Fig.\il(\ref{Fig:Plotsigmalimit3W}).
In Sec.\il(\ref{Sec:extreme-inversion}), the inversion  radii and inversion  plane extremes and limits are discussed.
We  focus on the necessary conditions for the existence of the inversion points of the counter-rotating and co-rotating flows in Sec.\il(\ref{Sec:all-to-rediid}), considering first the condition $\ell=$constant and then $\{\ell,\Em,\La\}$ constant. 
\subsubsection{Extremes and limits of the flows inversion points}\label{Sec:extreme-inversion}
We consider functions $r_\Ta^\pm$ in all generality, considering $\ell$ constant,  but not taking into account the condition of normalization or the conditions on the signs of $(\Em,\La)$.
There is
\bea\label{Eq:ci-ess-n-vorr-limit-sigmt}
\lim_{\sigma_\Ta\rightarrow0}r_{\Ta}^{\pm}=r_{\pm},\quad
\lim_{\ell\rightarrow (\pm \infty)} r_{\Ta}^{\pm}=r^\pm_{\epsilon}\quad\mbox{respectively and}\quad\lim_{a\rightarrow +\infty} \sigma_\Ta=1,\quad
\lim_{\ell\rightarrow \pm \infty} \sigma_\Ta=\sigma_{erg}.
\eea
The last limit is well defined (i.e.  $\sigma_\Ta\in [\,0,1\,]$), only for  $r_\Ta\leq 2$--Figs\il(\ref{Fig:Plotsigmalimitover}).
\begin{figure}
\centering
    \includegraphics[width=8cm]{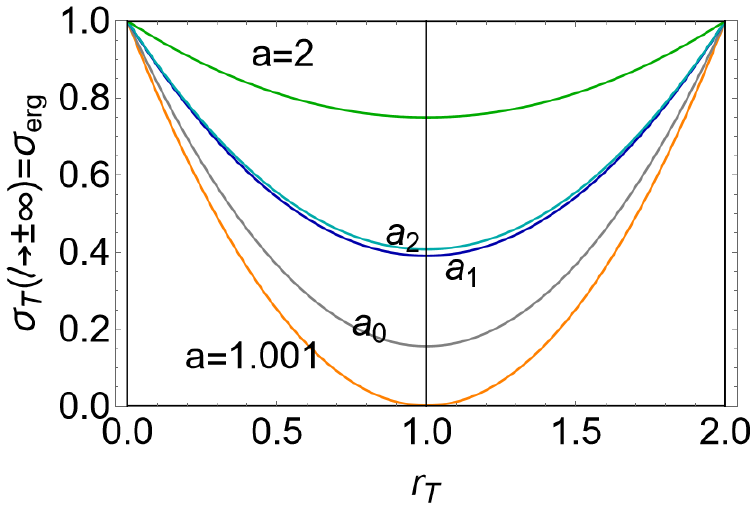}
       \caption{Analysis of the inversion point plane $\sigma_\Ta$ considered in Sec.\il(\ref{Sec:inversionsigmacontro}). Here the most general solution $\sigma_\Ta$ is  shown, not considering  the further constraints of Sec.\il(\ref{Sec:all-to-rediid})). Limit for large specific angular momentum magnitude, defined in Eq.\il(\ref{Eq:ci-ess-n-vorr-limit-sigmt}),  coincident with the ergosurface $\sigma_{erg}$  defined in Eqs.\il(\ref{Eq:sigma-erg}), for different {NSs} spin-mass ratios signed on the curves.  Spins $\{a_0,a_2\}$ are defined in Eqs\il(\ref{Eq:spins-a0-a1-a2}) and  $a_1 \equiv 1.28112:
\bar {r} _ {ms}^ -= \tilde {r} _ {ms}^-$ is defined in Sec.\il(\ref{Sec:extended-geo-struc}).  (All quantities are dimensionless).
  }\label{Fig:Plotsigmalimitover}
\end{figure}
The  limit of  the inversion radii on the poles is well defined  in the {BH} case only.  The limit for  large specific angular momentum in magnitude  is independent from the co-rotation or counter-rotation flow direction.
As noted also for the {BH} case, in the limit for large magnitude of the specific angular momentum, the inversion point approaches    the ergosurface, from above or  from below,  if the flow is counter-rotating or co-rotating  (this issue  is detailed  in Secs\il(\ref{Sec:all-to-rediid})).
A very large $\ell$ in magnitude is  typical of proto-jets emission or  quiescent   toroids orbiting  far from the central attractor for counter-rotating tori (with $\ell=\ell^+<0$) or some co-rotating tori with $\ell=\ell^->0$) in a class of slow rotating {NSs}.
The limit of  $r_\Ta=r_\Ta^{\pm}$ for  $ a\rightarrow+\infty$ (a very faster spinning {NS})  is not well defined. Increasing  the   spin $a>0$,  the inversion point approaches  the equatorial plane $(\sigma_\Ta\leq 1)$--Figs\il(\ref{Fig:Plotsigmalimitover},\ref{Fig:Plotsigmalimit},\ref{Fig:Plotvinthallo},\ref{Fig:Plotvinthalloa}).  It is interesting to note  that, contrary  to the {BH} case, in the {NS}  there can be  two inversion points radii $r_\Ta^\pm$.
\begin{figure}
\centering
    \includegraphics[width=8cm]{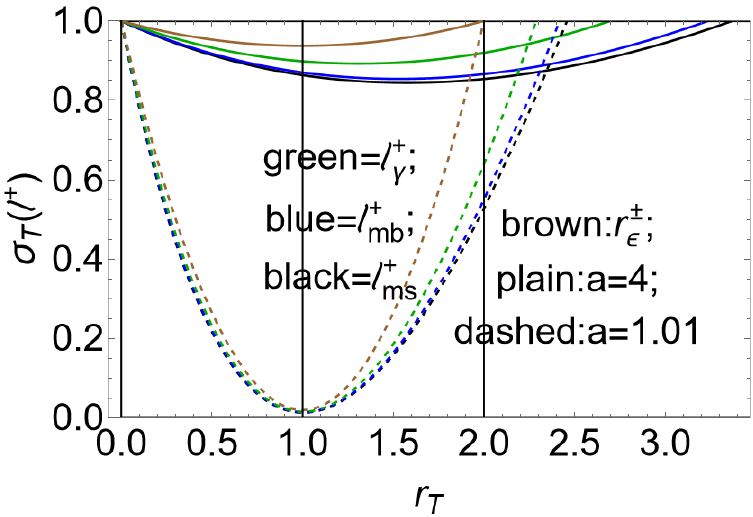}
      \includegraphics[width=8cm]{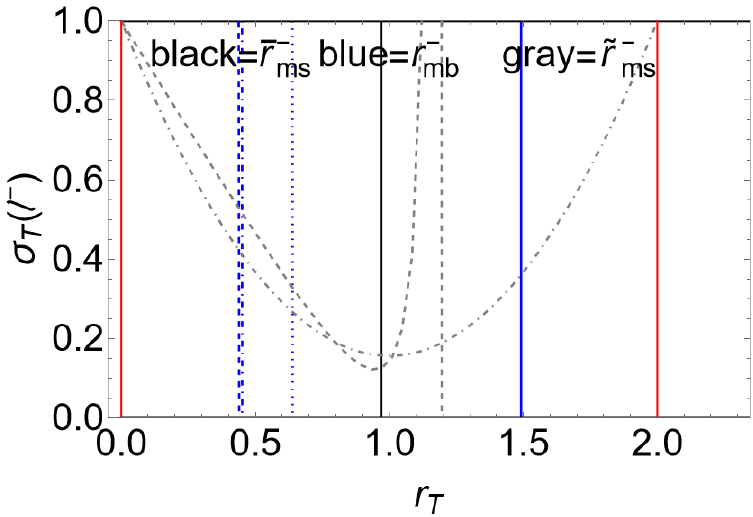}
       \caption{Inversion point plane $\sigma_\Ta$ considered in Sec.\il(\ref{Sec:inversionsigmacontro})  evaluated on the specific angular momentum  $\ell^+$ (left panel) and  $\ell^-$ (right panel)  as function of the inversion radius $r_\Ta$. Here the most general solution $\sigma_\Ta$ is  shown, not considering  the further constraints of Sec.\il(\ref{Sec:all-to-rediid})). (All quantities are dimensionless). There is  $\sigma\equiv\sin^2\theta\in[\,0,1\,]$, and $\sigma=1$ is the equatorial plane.   Left panel: counter-rotating inversion planes for counter-rotating flows $\ell=\ell^+<0$  evaluated at specific angular momentum on the marginally stable orbit $(ms)$, marginally bounded orbit $(mb)$  and marginally circular orbit $r_\gamma^+$, for different spin-mass ratios, as functions of the radius inversion point.
       Radius $r_{ms}^-=\{\bar{r}_{ms}^-,\tilde{r}_{ms}^-\}$ is the marginally stable orbit, $r_{mb}^-$ is the  marginally bounded orbit, for  $\ell=\ell^-$.    Right panel:
       Analysis of the inversion point radius $r_\Ta$  and plane $\sigma_\Ta$  from accretion flows with specific angular momentum $\ell=\ell^-$.  Dashed curve is for $a = 1.01$; dotted -
 dashed  curve is for $a=a_0$, dotted  curve is for $a=a_2$, plain curve for  $a = 2$. Spins $\{a_0,a_2\}$ are defined in Eqs\il(\ref{Eq:spins-a0-a1-a2}).}\label{Fig:Plotsigmalimit}
\end{figure}

{We proceed below as follows: we  first consider the inversion surfaces morphology analysing the critical radius $r_{cr}$, defined as $r_{cr}=r_\Ta^+=r_\Ta^-$, which  provides indication on the inversion surface radial structure, as the maximum  radial distance from the central spinning attractor, according to the  NS spin  and flow parameters.  Secondly,  we concentrate our attention  on the inversion points  in the regions close  to the singularity poles. The inversion surfaces  is in fact  in some cases not well defined in the region. On the other hand, in other cases the particular structure of the surfaces close  the poles singles  out this special region of the spacetime by the effects associated to the presence of the inversion points--see Fig.\il(\ref{Fig:Plotsigmalimit3W}. In this framework, it could be expected that this region may be of interest  for jet emission (having funnels along the rotational  axis) or other flows coming towards the singularity poles on the vertical direction.  However, most of the accretion flows of  the common scenarios currently under scrutiny concerns the singularity equatorial plane. For example,  accretion from a disk inner edge is usually  considered from the inner part of the disk laid  on the attractor (and disk) equatorial plane. Hence, we will  give particular attention to the inversion surfaces on the equatorial plane. This part of our analysis is then finalized by the discussion of the surfaces extreme points. By representing the inversion surfaces in the plane $(r,\sigma)$, we start  examining the extremes of the inversion plane $\sigma_\Ta$ according to the flow parameter $\ell$,  or the NS spin $a$ and finally to the inversion radius $r_\Ta$. In this way we closely relate   the inversion surfaces to the flows properties, considering  variation  of the inversion plane with the respect to the flow  specific angular momentum at different $a$ and radius $r_\Ta$.
We will show also that there is in fact an extreme of the  poloidal angle at the inversion point as function of the central singularity spin, showing that  the inversion surfaces can show strongly different features for different attractors.
Fixing $a$ and flow momentum $\ell$, we proceed considering  a general inversion surface expressed as the function $\sigma_\Ta(r_\Ta)$, tracing its profiles examining  the zeros of the function $\partial_{r_\Ta} \sigma_\Ta$. Results of this analysis are also confirmed by the  examination of the maxima  and minima of the  inversion radius function $r_\Ta$, with respect to the spin $a$, the flow  $\ell$, providing the maximum and minimum  radial distance of the inversion  point  from  the central singularity, with the respect to the NS spin or the parameter $\ell$. This study  ultimately fixes the vertical and radial extension of the inversion surfaces  which are spacetime geometrical properties defined as  toroidal surfaces embedding the central  singularity (ergoregions) presenting  a complex structure in the region close to the attractor poles and with a non--trivial variation with the spin of the singularity.
}

\medskip

\textbf{On the boundary  radius $r_\Ta^\pm=r_{cr}$}

The inversion radius $r_\Ta(\sigma_\Ta)$ depends on the geometry and torus parameter $(a,\ell)$, while  $r_{cr}$,
of Eqs\il(\ref{Eq:rcorss-defs})   is   a background property,
 independent  explicitly from  $\ell$. It represents a boundary value for   $r_\Ta^{\pm}$--see Figs\il(\ref{Fig:Plotvinthallo},\ref{Fig:Plotvinthalloa},\ref{Fig:plotrcross}).
 Radius $r_\Ta(\sigma_\Ta)$ provides  a good indication of the maximum extension of the inversion point for  $\sigma$ and for  $a$, as shown for the counter-rotating case  in Figs\il(\ref{Fig:Plotvinthallo}),
Figs\il(\ref{Fig:Plotvinthalloa}). From  Figs\il(\ref{Fig:plotrcross}), we can note that  the inversion radius $r_\Ta^{\pm}=r_{cr}$ always increases with the {NS} spin and decreases  with the plane $\sigma$, being larger at the poles--Fig.\il(\ref{Fig:Plotsigmalimit3W}).  Function $r_\Ta^\pm=r_{cr}$ has no extremes for plane $\sigma$. There is however
$\partial_a r_ {cr} = 0 $ on the equatorial plane.
Below we study  the maximum geometric extension of the inversion spheres on the equatorial plane and on the vertical direction parallel to the central attractor rotational axis, which is a  relevant  aspect particularly  for  proto-jets flows.

\medskip

\textbf{Inversion points close to the  poles and on the equatorial plane}

We  concentrate  on the inversion points close to   the {NS} poles $\sigma=0$ and on the  {NS}  equatorial plane $\sigma=1$. (The inversion surfaces on the equatorial plane are shown in Fig.\il(\ref{Fig:Plotsigmalimit3W})).
      There is
\bea&&\label{Eq:coord-vist-freq}
\sigma_\Ta = 0:\ \mbox{for}\quad (\ell = 0,r_\Ta > 0)\quad\mbox{and}
\quad  \sigma_\Ta = 1:\ \mbox{for}\quad \ell <0\, \cup\,  \ell \geq \ell_\gamma^-,\quad r_\Ta=r_b,\quad\mbox{where}\quad
      r_b\equiv \frac{2(\ell- a)}{\ell }.
\eea
In particular there is $\sigma_\Ta=1$ in $r_\Ta >2$ only for  $a>1, \ell <0, r_\Ta =r_b$.
In other words, on the equatorial plane ($\sigma_\Ta=1$) at  inversion point, there is
\bea\label{Eq:lbeta}
 \ell_\beta\equiv  \frac{2 a}{2-r},\quad\mbox{or}\quad r_\Ta=r_b=\frac{2(\ell-a)}{\ell },
\eea
 for $\ell<0$ (counter-rotating flows) or for   $\ell>\ell_\gamma^->0$, whereas there are no solutions for $\ell_\gamma^->\ell>0$.

\medskip

\textbf{Extremes of the inversion  plane $\sigma_\Ta$}

Below we analyze the extremes of the inversion plane $\sigma_\Ta$ with respect to the inversion point radius $r_\Ta$, the fluid specific angular momentum $\ell$ and  the dimensionless {NS} spin $a$.

There is
\bea\nonumber
&&\partial_\ell \sigma_\Ta=0:\quad\mbox{for}\quad (\ell \neq0, r_\Ta=0)
\\\nonumber
&&
\partial_a \sigma_\Ta=0:\quad\mbox{for}\quad (\ell <0, r_\Ta=\{0,r_{\Ta\max}^0\}), (\ell =0, r_\Ta>0), (\ell >0, r_\Ta=0),
\\\nonumber
&&
\quad\quad
 \quad  \mbox{where}\quad r_{\Ta\max}^0\equiv \frac{1}{2} \left[a \ell +2\sqrt{a^2 \left(\ell ^2+4\right)+4(1- a \ell)}\right].
\\\nonumber
&&
\partial_{r_\Ta}\sigma_\Ta=0:\quad\mbox{for}\quad  (\ell <0, r_\Ta=r_{\Ta\max}^+), (\ell =0, r_\Ta>0), (\ell =\ell_\gamma^-, r_\Ta=0), (\ell >\ell_\gamma^-, r_\Ta=r_{\Ta\max}^-),
\\
&&\label{Eq:rmax}
\quad\mbox{with }\quad
r_{\Ta\max}^\mp\equiv \frac{1}{2} \left[a \ell\mp\sqrt{a \left[a \left(\ell ^2+4\right)-4 \ell \right]}\right],\quad\mbox{where}\quad  \sigma_{\Ta\max}^\pm\equiv \sigma_{\Ta}(r_{\Ta\max}^\pm)
\eea
--Figs\il(\ref{Fig:Plotsigmamax}).
\begin{figure}
\centering
    \includegraphics[width=8cm]{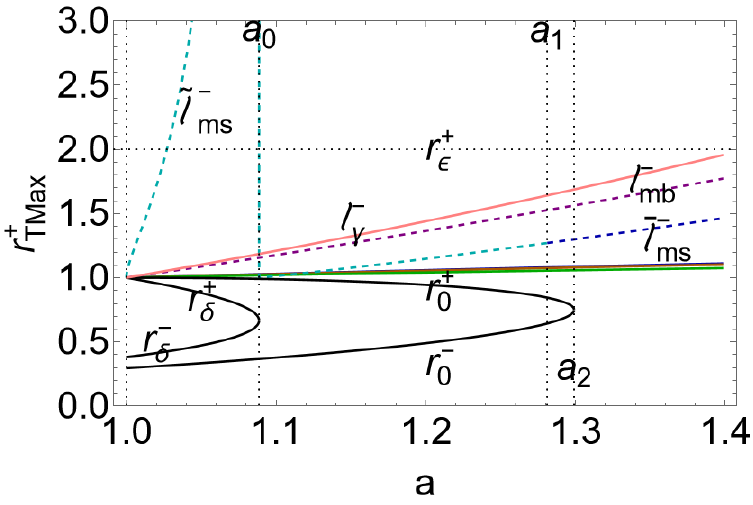}
      \includegraphics[width=8cm]{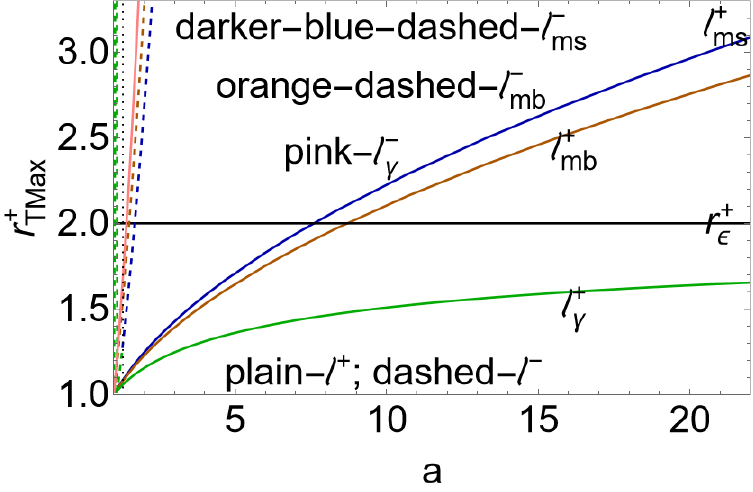}
      \includegraphics[width=8cm]{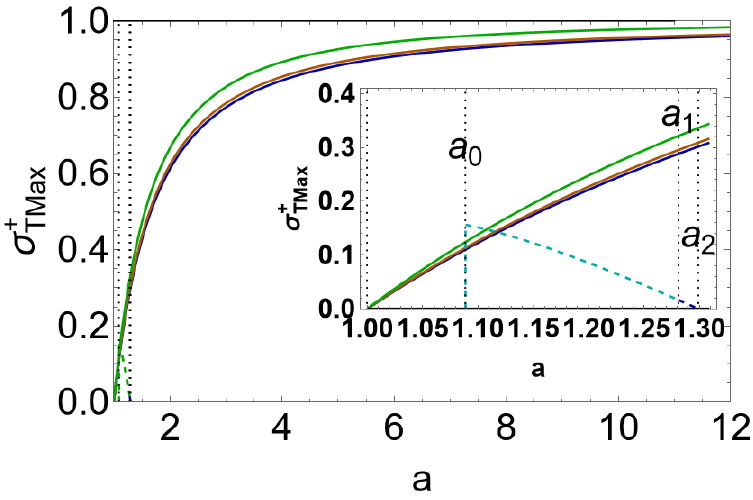}
     \includegraphics[width=8cm]{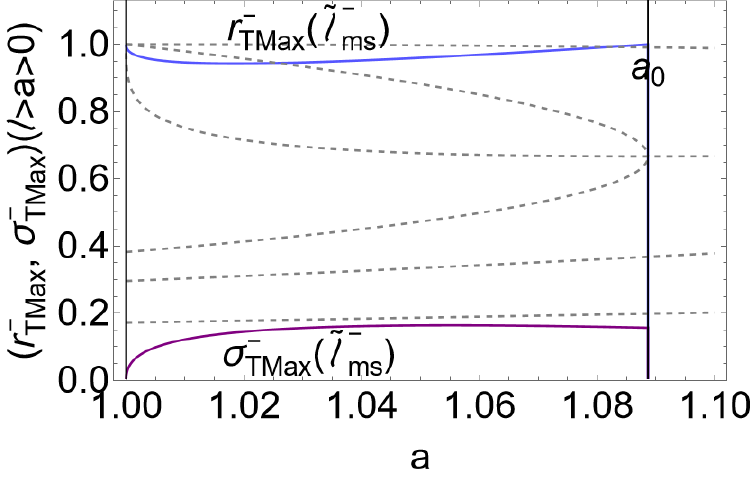}
  \caption{Analysis of the extremes of the flow inversion point plane  $r_{\Ta\max}^{\pm}: \partial_{r_\Ta}\sigma_\Ta=0$ of Eqs\il(\ref{Eq:merka.til-vant-sal-qua-maxmin}) and maximum inversion plane $\sigma_{\Ta\max}^+$ of Eqs\il(\ref{Eq:rmax}).   Spins $\{a_0,a_2\}$ are defined in Eqs\il(\ref{Eq:spins-a0-a1-a2}) and  $a_1 \equiv 1.28112:
\bar {r} _ {ms}^ -= \tilde {r} _ {ms}^-$ is defined in Sec.\il(\ref{Sec:extended-geo-struc}).  See also analysis of  Secs\il(\ref{Sec:inversionsigmacontro}). (All quantities are dimensionless). Upper line panels: radius $r_{\max}^{\pm}$  evaluated on the specific angular momenta $\ell^+<0$ and $\ell^-$ on the marginally bounded orbit $(mb)$, marginally stable orbits $(ms)$-- Figs\il(\ref{Fig:Plotdicsprosc}). Radius $r_\epsilon^+=2$ is the outer ergosurface on the equatorial plane. Upper left panel is a close--up  view of the right panel.
  Bottom left panel: the maximum plane $\sigma_{\Ta\max}^+\equiv \sigma_{\Ta}(r_{\Ta\max}^+)$ as function of the {NS} spin--mass ratio, for different fluid specific angular momenta following the  colouring notation of the upper right panel. Inside plot is a close-up view. Bottom right panel:  radius $r_{\Ta\max}^-$ and plane $\sigma_{\Ta\max}^-\equiv \sigma_{\Ta}(r_{\Ta\max}^-) $ on the fluid specific angular momentum $\tilde{\ell}_{ms}^-\equiv \ell^-(\tilde{r}_{ms}^-)>a>0$. Dashed-gray curves are the Kerr {NS}  geodesic structure. (In this analysis  the most general solutions $r_\Ta^\pm$ are shown not considering    the constants of motion  analysis of  Sec.\il(\ref{Sec:all-to-rediid}).)).}\label{Fig:Plotsigmamax}
\end{figure}
{
Therefore the  inversion surface morphology  (plane $\sigma_\Ta$), for photons and particles, differentiates the central attractors by their  spin, and the extreme points of the spheres depend on the spin and fluid parameter $\ell$, distinguishing proto-jets from accretion tori.}

\textbf{Extremes of the inversion point radius $r^\pm_\Ta$}

Radius $r_{cr}$ always decreases with $\sigma\in\, ]\,0,1\,[$--see Figs\il(\ref{Fig:Plotvinthallo},\ref{Fig:Plotvinthalloa},\ref{Fig:plotrcross}). Whereas  there is
 $\partial_a r_ {cr} = 0$, for the limiting case $\sigma=1$ (the equatorial plane), increasing always with the spin for $\sigma\in\,]\,0,1\,[$.
 There is $r_ {cr}(\sigma \rightarrow
    1) = 0$.

   The extremes of the inversion radii $r_\Ta^\pm$ for the dimensionless {NS} spin $a$ are  as follows
\bea\label{Eq:merka.til-vant-sal-qua-maxmin}
&&
\partial_a r_\Ta^-=0:\quad \sigma_\Ta =1, (\ell <0, \ell >\ell_\gamma^-),\quad\mbox{and}\quad
\partial_a r_\Ta^+=0: (\sigma_\Ta \in \,]\,0,1\,[\, , \ell =\ell_e^-<0), (\sigma_\Ta  =1, \ell \in \,]\,0,\ell_\gamma^-\,[  \, ),
\\\label{Eq:lemp}
&&
\mbox{where} \quad
%\ell_e^\mp\equiv\frac{\sigma_\Ta  }{a (\sigma_\Ta  -1)}-\sqrt{\frac{\sigma_\Ta  ^2 \left[1-a^2 (\sigma_\Ta  -1)\right]}{a^2 (\sigma_\Ta  -1)^2}};\quad
 \ell_e^-\equiv\frac{a \sigma_\Ta \left[1+\sqrt{1+r_{cr}^2}\right] }{-r_{cr}^2}
\eea
see--Figs\il(\ref{Fig:Plotsigmalimit3}) and also Figs\il(\ref{Fig:Plotsigmalimit3W}).
\begin{figure}
\centering
    \includegraphics[width=8cm]{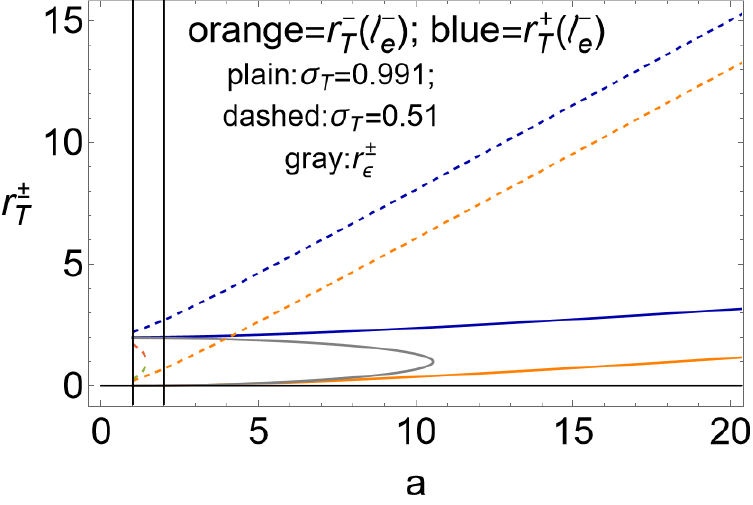}
    \includegraphics[width=8cm]{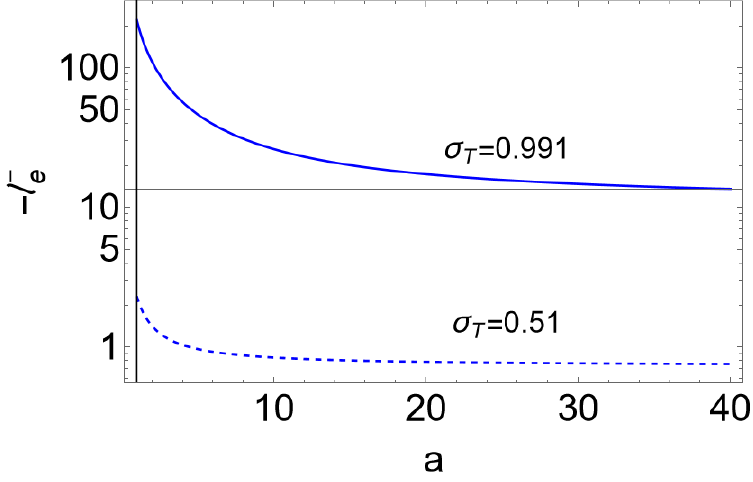}
       \caption{Analysis of the extremes of the flow inversion point radius $r_\Ta^{\pm}$ of Eqs\il(\ref{Eq:merka.til-vant-sal-qua-maxmin}). (All quantities are dimensionless).  Right  panel: fluid specific angular momentum $\ell_e^-$  defined in Eqs\il(\ref{Eq:lemp}), as functions of the spin-mass ratio $a$, for different planes $\sigma\equiv\sin^2\theta$ (where $\sigma=1$ is the equatorial plane). Left panel: inversion   radii   $r_\Ta^{\pm}$ on $\ell_e^-$ as functions of $a$, for different $\sigma$. Radii $r_{\epsilon}^+>r_{\epsilon}^-$ are the outer and inner ergosurfaces respectively.}\label{Fig:Plotsigmalimit3}
\end{figure}

(There are no solutions  $\partial_{\sigma_\Ta} r_\Ta^{\pm}=0$, and
$\partial_a r_\Ta^+=0$  for  $\ell=0$--Figs\il(\ref{Fig:Plotvinthalloa}))

Considering the  inversion point radius as function of the fluid specific angular momentum $\ell$, there is
\bea&&\label{Eq:istr-stat-siran}
\partial_\ell r_\Ta^-=0:\quad \sigma_\Ta =1\quad (\ell <0,\ell>\ell_\gamma^->1);\quad
\partial_\ell r_\Ta^+=0:\quad \sigma_\Ta  =1\quad(\ell\in\,]\,0,1\,]\, ,\ell_\gamma^->\ell>1),
\eea
(see also Eqs\il(\ref{Eq:coord-vist-freq},\ref{Eq:lbeta})).
There are  extremes  of the inversion point radius, as function of $\ell$ on the equatorial plane.

For the co-rotating flows inversion points ($\ell>0$), as in Eqs\il(\ref{Eq:istr-stat-siran}), (\ref{Eq:merka.til-vant-sal-qua-maxmin}), (\ref{Eq:rmax}), (\ref{Eq:coord-vist-freq}),  condition  $\ell>\ell_\gamma^->1$, has to be evaluated. (There is $\ell_\gamma^-=a$, for $\ell^-$ evaluated at the last circular co-rotating orbit $r_\gamma^-=0$ and on $r_{[\gamma]}^-=a^2$, bottom boundary for the $\mathbf{L_3^-}$ range.).
{
There is:
\bea\label{Eq:condizioni-lla-rdelts1}
&&
\ell^-=\ell_\gamma^-:\quad\mbox{for}\quad \left(r=r_{[\gamma]}^-, a\neq a_3\right), (r=0, a\geq 1),\quad \mbox{where}\quad a_3\equiv \sqrt{3},
\\\nonumber
&&
\ell^->\ell_\gamma^-:\quad\mbox{for}\quad
a\in\, [\,1,a_0\,[\, : \left(r\in\, ]\,r_{\delta }^-,r_\delta^+\,[\, , r>r_{[\gamma]}^-,r\neq r_\delta^{(1)}\right),
\\&&\nonumber
\;
\qquad\qquad\quad\mbox{for}\quad a\in \,[\,a_0,a_3\,[\, : \left(r>r_{[\gamma]}^-, r\neq r_\delta^{(1)}\right),\;
\\&&\nonumber
\;
\qquad\qquad\quad\mbox{for}\quad
a\geq a_3: r>r_{[\gamma]}^-,
 \eea
 where
 \bea
&&
%
%\bea
\label{Eq:rdelta1}
r_\delta^{(1)}\equiv\frac{\delta^++\delta^-+8}{6},\quad \mbox{and}\quad\delta^\pm\equiv 2^{2/3} \sqrt[3]{27 a^2\pm3 \sqrt{81 a^4-96 a^2}-16},
\eea
there is $\ell_\gamma^-=a$ and $r_{[\gamma]}^-=a^2$--Figs\il(\ref{Fig:delta-plot-split}).
}
\begin{figure}
\centering
    \includegraphics[width=8cm]{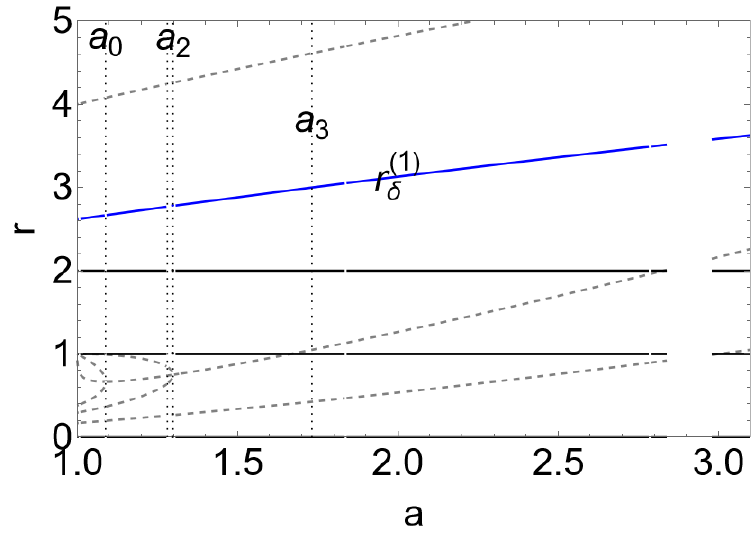}
       \caption{Radius $r_\delta^{(1)}$, defined in Eq\il(\ref{Eq:rdelta1}), as function of the {NS} spin-mass ratio.  It refers to the conditions $\ell^-\geq a$ of Eqs\il(\ref{Eq:condizioni-lla-rdelts1}). Gray dashed curves are the geodesic structure of the {NSs} as in Figs\il(\ref{Fig:Plotdicsprosc}). There is $a_3=\sqrt{3}$. Spins $\{a_0,a_2\}$ are defined in Eqs\il(\ref{Eq:spins-a0-a1-a2}) and  $a_1 \equiv 1.28112$. (All quantities are dimensionless).
  }\label{Fig:delta-plot-split}
\end{figure}
 \begin{figure}
\centering
       \includegraphics[width=5.75cm]{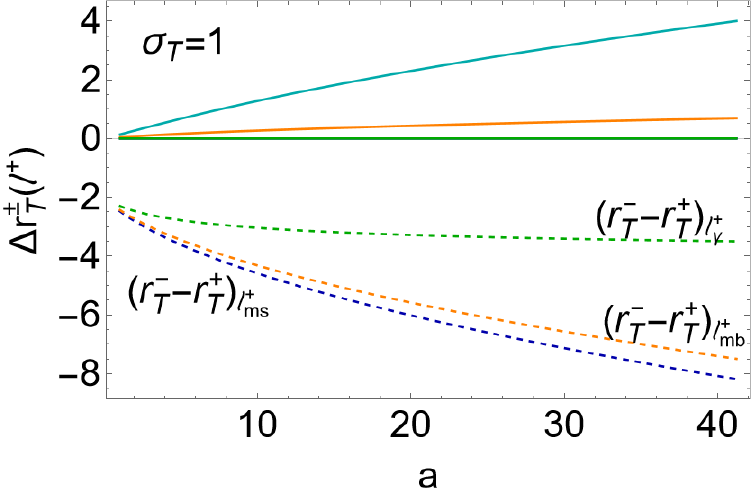}
    \includegraphics[width=5.75cm]{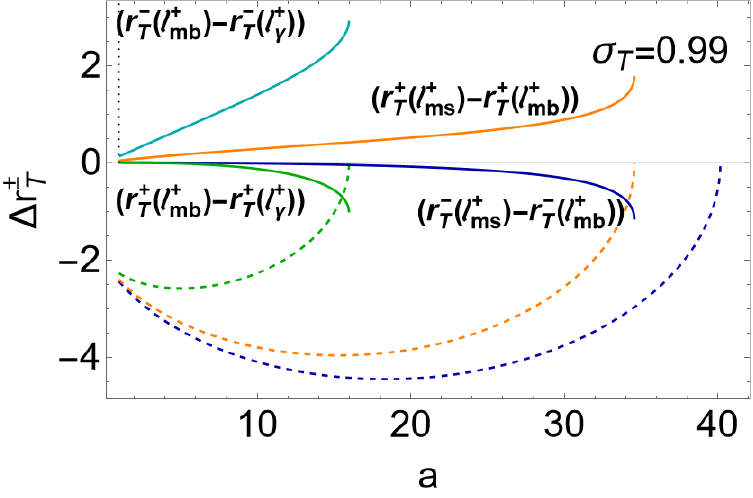}
    \includegraphics[width=5.75cm]{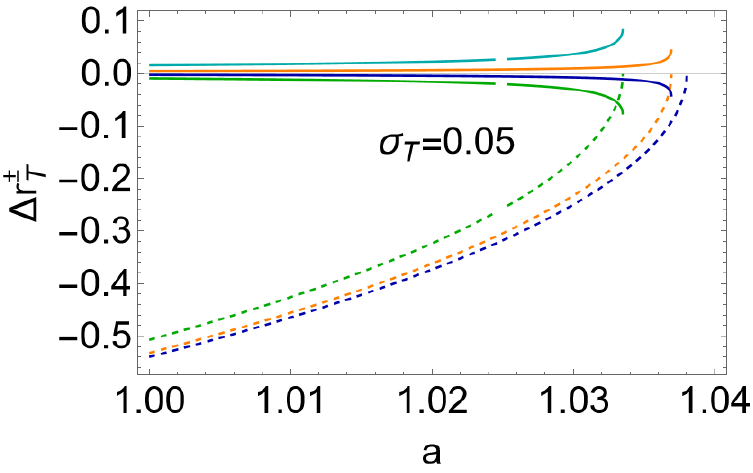}
  \caption{Analysis of the counter-rotating flow inversion radius  $r_\Ta^{\pm}$ of Eqs\il(\ref{Eq:rinversionmp})  with specific angular momentum $\ell^+<0$, as functions of the {NS} spin-mass ratio $a$, for different planes $\sigma\equiv \sin^2\theta$ signed on the panels, where $\sigma=1$ is the equatorial plane.  (Here most general solutions $r_\Ta^\pm$ shown not considering    constraints of Sec.\il(\ref{Sec:all-to-rediid}).) Difference $\Delta r_\Ta^{\pm}$ defined on the curves according to the notations of the left  and center panels. Inversion points are evaluated at specific angular momentum on the marginally bounded orbit $(mb)$, marginally stable orbit $(ms)$ and marginally circular orbit $r_\gamma^+$. Dashed curves are differences $(r_\Ta^--r_\Ta^+)_\ell$ at fixed $\ell$. Plain curves are differences  $(r_\Ta^{\pm}(\ell_1)-r_\Ta^{\pm}(\ell_2))$
for two fixed values of $\ell$. (All quantities are dimensionless).}\label{Fig:Plotvinthalld}
\end{figure}
 \begin{figure}
\centering
      \includegraphics[width=8cm]{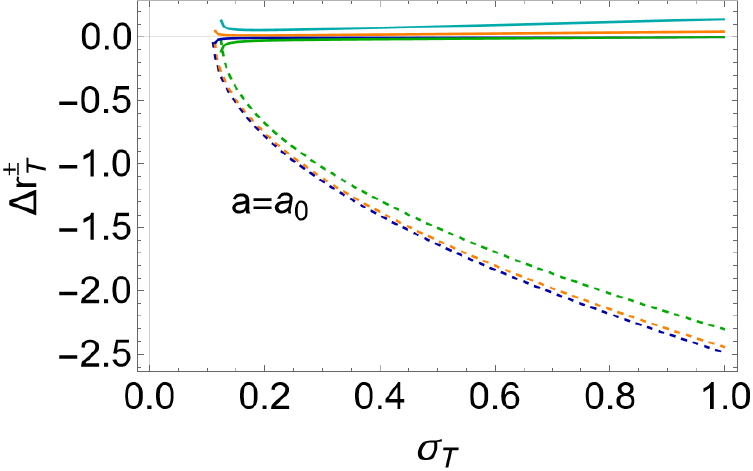}
                   \includegraphics[width=8cm]{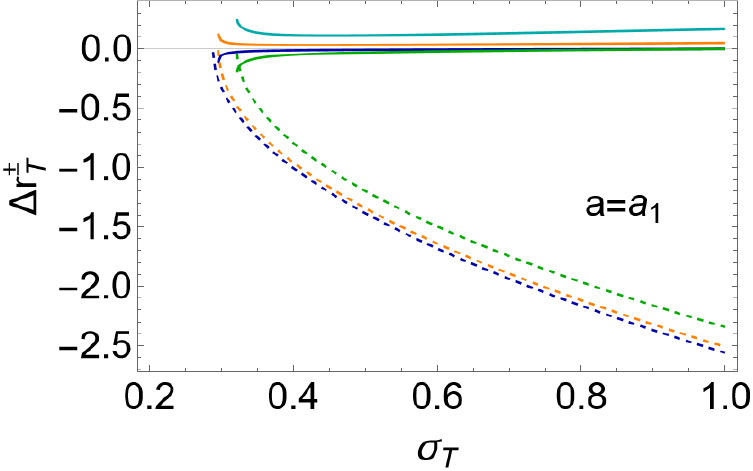}
                                    \includegraphics[width=8cm]{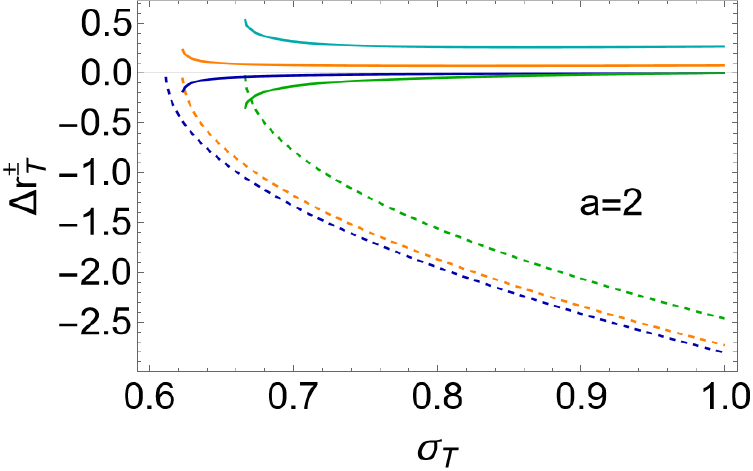}
       \includegraphics[width=8cm]{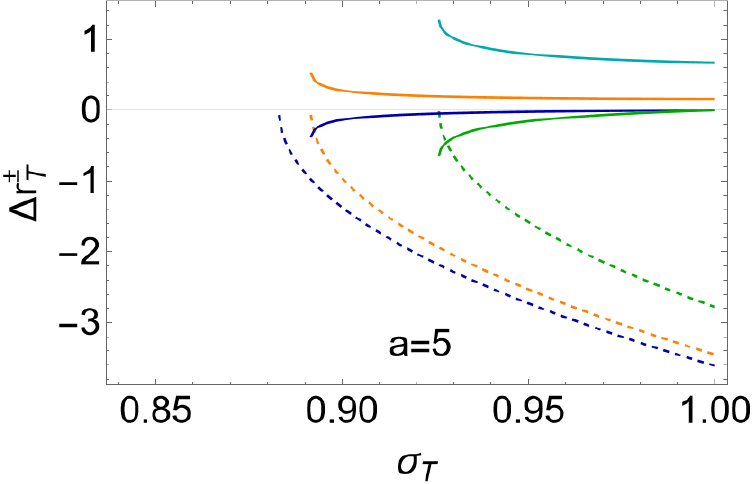}
  \caption{Analysis of the counter-rotating flow inversion radius  $r_\Ta^{\pm}$ of Eqs\il(\ref{Eq:rinversionmp})  with specific angular momentum $\ell^+<0$, as function of planes $\sigma_\Ta=\sigma\equiv \sin^2\theta\in[\,0,\,1]$, where $\sigma=1$ is the equatorial plane,  for different {NS} spin-mass ratios signed on the panel.  (Here most general solutions $r_\Ta^\pm$ shown not considering    constraints of Sec.\il(\ref{Sec:all-to-rediid}).) Spins $\{a_0,a_1\}$ are defined in Eqs\il(\ref{Eq:spins-a0-a1-a2}).  Difference $\Delta r_\Ta^{\pm}$ defined on the curves according to the notations of the left  and center panels of Figs\il(\ref{Fig:Plotvinthalld}) . Inversion points are evaluated at specific angular momentum on the marginally bounded orbit $(mb)$, marginally stable orbit $(ms)$ and marginally circular orbit $r_\gamma^+$. Dashed curves are differences $(r_\Ta^--r_\Ta^+)_\ell$ at fixed $\ell$. Plain curves are differences  $(r_\Ta^{\pm}(\ell_1)-r_\Ta^{\pm}(\ell_2))$
for two fixed values of $\ell$. (All quantities are dimensionless).}\label{Fig:PlotvinthalldW}
\end{figure}
There are no inversion points for $\ell=\ell^->0$ for $a>a_2$.
{The inversion corona  therefore is  a closed  region, generally of small extension and thickness.    The vertical coordinate $z_\Ta$, elongation  on the central rotational axis of the inversion point, for the counter-rotating flows, is the order of $\lesssim 1.4 $ (central attractor mass, where $r=\sqrt{z^2+y^2}$ and $\sigma\equiv\sin^2\theta=y^2/(z^2+y^2)$.).}
\section{Inversion points of the counter-rotating and co-rotating flows}\label{Sec:all-to-rediid}
  Inversion point  definition, as the set of points where $u^\phi=0$ identifies a surface, inversion surface, surrounding the central singularity,
depending only on $\ell$ parameter of the freely infalling (or outfalling) matter. The  inversion surface is a general property of the orbits in the Kerr  {NS} spacetime. In this article we  study   in  general  function $\ell(u^\phi=0)$, constraining   then $\ell_\Ta$ when framed in the particles flow and tori  flows inversion points. We assume:  \textbf{1.}  constance of  $(\Em_\Ta, \La_\Ta)$  (implied by  $\ell=$constant) evaluated
at the inversion point with $\dot{t}$>0.
\textbf{2.} We then consider the  normalization condition at the  inversion point,  distinguishing particles and photons in the flow.
 \textbf{3.} The matter flow is then related to the orbiting structures,  constraining  the range  of values for  $\ell$, to  define the inversion corona for proto-jets or accretion driven flows, as background geometry properties, depending only on the spacetime spin.
(Inversion surface  describes also  particles  with $\dot{r}_\Ta>0$  (outgoing particles) {or particles with an axial velocity $\dot{\theta}\neq0$ along the {BH} rotational axis.}).

Sign of $\{\ell,\La\}$ at the inversion point for the  counter-rotation case   is addressed in Sec.\il(\ref{Sec:sever-amernothermed}). 
In this section we explore in details the inversion points of the accreting flows.  Specifically,  counter-rotating flows $(\ell<0)$ inversion points  are detailed in Sec.\il(\ref{Sec:inversion-counter-rot}). In
 Sec.\il(\ref{Sec:counter-rotating-inversion-NSflow}) we analyze  the inversion radius  $r_\Ta^{\pm}$,  while in  Sec.\il(\ref{Sec:inversionsigmacontro}) we focus on the
inversion plane $\sigma_\Ta$.
The co-rotating  ($\ell>0$) flows inversion points  are detailed in Sec.\il(\ref{Sec:corot-inversion}).
\begin{figure}
\includegraphics[width=5.75cm]{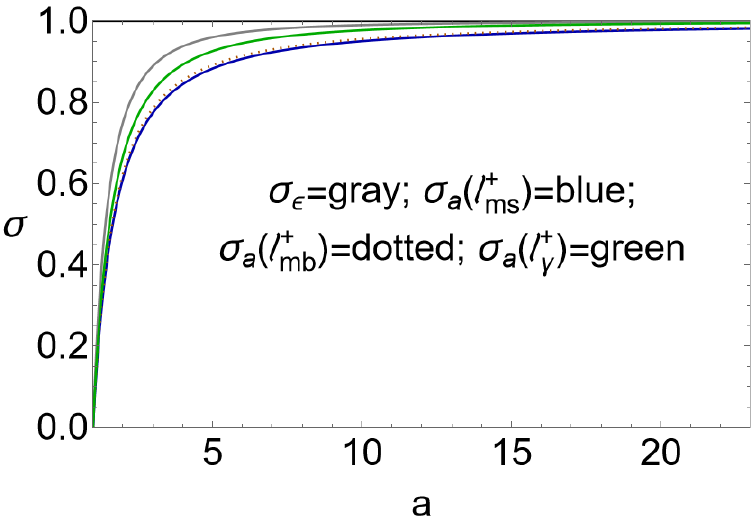}
 \includegraphics[width=5.75cm]{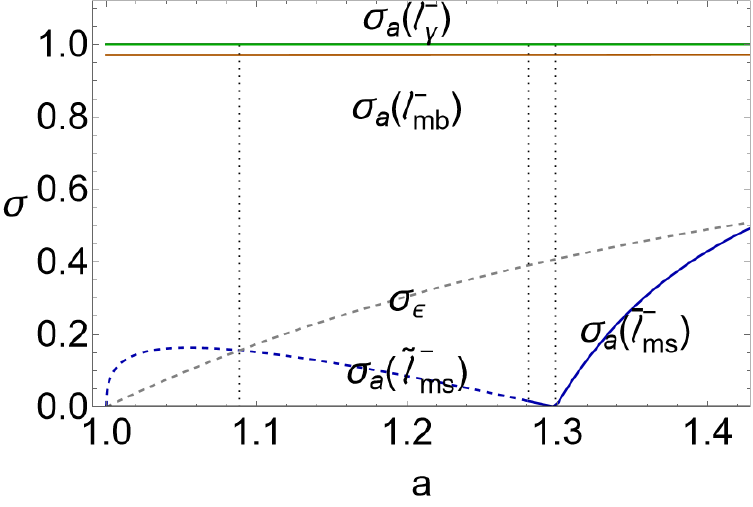}
                     \includegraphics[width=5.75cm]{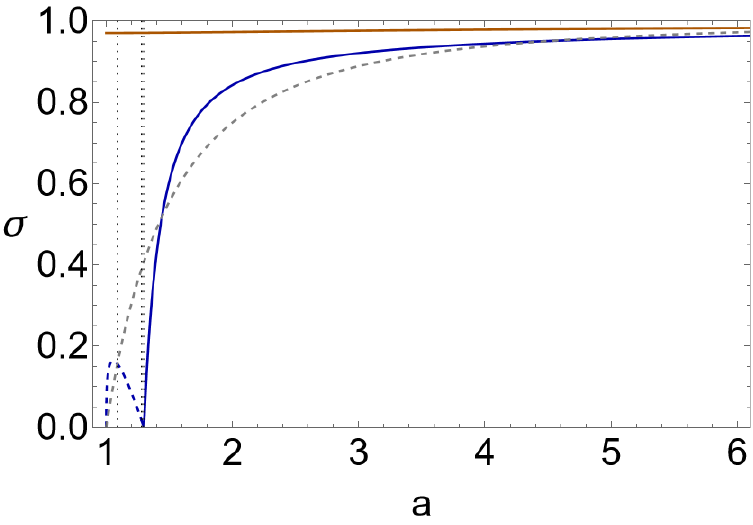}
     \caption{Plane $\sigma_{a}$, defined in Eq.\il(\ref{Eq:sasW}), as function of the {NS} spin-mass ratio  $a>1$ evaluated on the counter-rotating fluid specific angular momentum $\ell^+<0$ (left panel). Center and right panels show the plane $\sigma_{a}$ for different specific angular momenta $\ell^-$, center panel is a close--up view of the right panel. Color notation follows the  closed-up view  panel.   Notation $(mb)$ is for quantities evaluated on the marginally bounded orbit, $(ms)$ for marginally stable orbit, $(\gamma)$ for marginally circular orbit.
      Plane $\sigma_\epsilon$ is defined  in Eq.\il(\ref{Eq:fioc-rumor}). The ergoregion is defined for $\sigma\in[\,\sigma_\epsilon,\,1]$. There is $\sigma\equiv \sin^2\theta\in[\,0,1\,]$, where $\sigma=1$ is the equatorial plane.
       Dotted vertical lines are spins $a_0<a_1<a_2$, defined in Eqs\il(\ref{Eq:spins-a0-a1-a2}). The marginally stable orbit is $r_{ms}^{-}\equiv\{\tilde{r}_{ms}^-,\bar{r}_{ms}^-\}$, respectively there is $\{\tilde{\ell}_{ms}^-,\bar{\ell}_{ms}^-\}$--see  Sec.\il(\ref{Sec:extended-geo-struc}).  (All quantities are dimensionless). } \label{Fig:Plotvinthallosrc12}
\end{figure}
\subsection{Inversion points of counter-rotating flows}\label{Sec:inversion-counter-rot}
In this section we study  the inversion points of counter-rotating flows $(\ell<0)$, considering in Sec.\il(\ref{Sec:counter-rotating-inversion-NSflow}) the inversion radius  $r_\Ta^{\pm}(\sigma)$, while in Sec.\il(\ref{Sec:inversionsigmacontro}) we explore the inversion plane   $\sigma_\Ta(r)$. In  Sec.\il(\ref{Sec:counter-rotating-inversion-NSflow})  and  Sec.\il(\ref{Sec:inversionsigmacontro}) we develop our analysis  considering  definitions in Eq.\il(\ref{Eq:lLE}), derived by the definition of $\ell=$constant and condition $u^\phi=0$.
Whereas in Sec.\il(\ref{Sec:sever-amernothermed}), we discuss  the conditions for $u^\phi=0$ and $\ell<0$,
 within $(\Em,\La)$ constant, considering  definitions in Eq.\il(\ref{Eq:lLE}).

According to the analysis in  Sec.\il(\ref{Sec:coro-contro}) and Sec.\il(\ref{Sec:extended-geo-struc})  we  consider \textbf{(I)} tori with $\ell=\ell^+<0$,  that is with $(\La<0,\Em>0)$,  located (centered) at $r>r_\gamma^+$-- Figs\il(\ref{Fig:Plotvinthallo},\ref{Fig:Plotvinthalloa},\ref{Fig:Plotvinthalld},\ref{Fig:PlotvinthalldW}). { \textbf{(II)}  Tori with momentum $\ell=\ell^-<0$, that is with $(\La<0,\Em>0)$, in
$]\,r_0^-,r_\delta^-\,[\,\cup\,]\,r_\delta^+,r_0^+\,[$, for spacetimes $a\in[\,1, \,a_0]$, and in the orbital region $]\,r_0^-,r_0^+\,[$, for {NSs} with spin $a\in\,]\,a_0,a_2\,]$.} The case  $\ell=0$ (that is on $r_{0}^\pm: \La=0$) is considered  in Sec.\il(\ref{Sec:inversionlequal0}). On $r_\delta^{\pm}$, there is $\Em=K=0$ and $\La<0$.
(There are no  inversion points for  $\La>0,\Em<0$.)

\subsubsection{Inversion radius  $r_\Ta^{\pm}$ of the counter-rotating $(\ell<0)$ flows}\label{Sec:counter-rotating-inversion-NSflow}
We concentrate on the existence of a inversion radius $r_\Ta^{\pm}$,  using the definitions Eq.\il(\ref{Eq:lLE}).
{For counter-rotating fluids $(\ell<0)$ there is
\bea&&\nonumber
\ell<0 \quad\mbox{for}:
 \\\label{Eq:div-d}
&&
 \sigma=\sigma_\Ta: \quad (r_\Ta\in\,]\,0,2\,]\, , \ell <0), (r_\Ta>2, \ell\in\,[\,\ell_\beta,0\,[\,)).
 \eea
Conditions (\ref{Eq:div-d})  can be expressed also as
\bea
\nonumber
&&\sigma_\Ta\in\,]\,0,\sigma_\epsilon\,[\, : (\ell =\ell_{a\sigma}, r=r_\Ta^-), (\ell\in\,]\,\ell_{a\sigma},0\,[\, , r_\Ta=r_\Ta^\pm); \quad \sigma_\Ta\in\,[\,\sigma_\epsilon,1\,[\, :  (r_\Ta=r_\Ta^\pm);\quad  \sigma_\Ta =1:  (r_\Ta=r_\Ta^+).
\eea
where
\bea
\ell_{a\sigma}^{\mp}\equiv\frac{a \sigma_\Ta }{1-r_{cr}^2}\mp\sqrt{\frac{a^2 r_{cr}^2 \sigma_\Ta ^2}{\left[1-r_{cr}^2\right]^2}},
\eea
%\\&&\nonumber
or  equivalently it is possible to express this condition  also as
\bea
&&\nonumber
\sigma_\Ta =\sigma_{a}: (r=r_\Ta^-); \quad \sigma_\Ta\in\,]\,\sigma_{a},1\,[\, : (r_\Ta=r_\Ta^\pm);\quad \sigma_\Ta=1: (r_\Ta=r_\Ta^+),
\eea
where
\bea\label{Eq:sasW}
\sigma_{a}\equiv\frac{1}{2} \left[\sqrt{\ell ^2 \left(-\frac{4 \ell }{a}+\ell ^2+4\right)}-\ell ^2\right]+\frac{\ell }{a},
\eea
(see Figs\il(\ref{Fig:Plotvinthallosrc12}) and Figs\il(\ref{Fig:Plotvinthallosrc12W})).}
There are no solutions on the poles i.e. $\sigma=0$
--Figs\il(\ref{Fig:Plotsigmalimit}).
There are two inversion points $r_\Ta^\pm(\sigma)$ for  fixed  $(\sigma_\Ta,\ell)$, while on the equatorial plane there is always one only radius $r_\Ta^+$. This implies that, at plane different from the equatorial, there are two radii, where  $\ell<0$ and  $u^\phi=0$. This fact, analyzed in details in Sec.\il(\ref{Sec:sever-amernothermed}),  constitutes a major  difference with the {BH} case.

The inversion point radius $r_\Ta^+$ on the equatorial plane is always \emph{out} of the ergoregion and there is
\bea\label{Eq:quatorial}
&&
 \sigma_\Ta =1: \ell <0, r_\Ta =r_\Ta^+\quad \mbox{equivalently}\quad
 r_\Ta>2, \ell =\ell_\beta,
\\&&\label{Eq:no-ergolneg54}
\mbox {there are \emph{no} solutions}\quad  \ell <
   0 \quad\mbox {for}\quad  r_\Ta\in [\,r_\epsilon^-, r_\epsilon^+\,]
   \eea
  --see also Figs\il(\ref{Fig:Plotvinthallo},\ref{Fig:Plotvinthalloa},\ref{Fig:Plotsigmalimit3},\ref{Fig:Plotsigmalimit3W}).
   More specifically, there are solutions
   {
   \bea\label{Eq:mon-gig-cin}
  &&
  r_\Ta = r_\Ta^-\quad
  \mbox{for}\quad r_\Ta\in \,]\, 0, r_ {\epsilon}^-\,[\, ,\sigma_\Ta\in\,[\,\sigma_{\epsilon}^+,1\,[ \\
  &&\nonumber \mbox{equivalently for}\quad \sigma =\sigma_\Ta:\quad
 r_\Ta\in\,]\,0,\,1[\, , \ell \leq \ell_\rho,\quad \mbox{where}\quad \ell_\rho\equiv-\frac{2 \left(a^2-1\right) r}{a (r-1)^2},
         \\ &&\nonumber
     r_\Ta = r_\Ta^+\quad    \mbox{for}\quad r_\Ta > r_\epsilon^+ ,\sigma_\Ta\in\, [\,\sigma_\epsilon,1\,]\, ,
     \\
     &&\nonumber \mbox{equivalently for } \quad \sigma =\sigma_\Ta: \quad(r_\Ta\in\, ]\,1,2\,]\, , \ell \leq \ell_\rho), (r_\Ta>2, \ell\in\,[\,\ell_\beta,\ell_{\rho}\,]\,).
\eea}
The counter--rotating case includes also tori with $\ell=\ell^+<0$ and  $\ell=\ell^-<0$ in the ergoregion for {NSs} with $a\in\,[\,1,a_2\,]$.
\subsubsection{Inversion plane $\sigma_\Ta$ of the counter-rotating  $(\ell<0)$ flows}\label{Sec:inversionsigmacontro}
We discuss the   necessary  conditions for the existence of a inversion plane $\sigma_\Ta\in[\,0,1\,]$   as function of  the inversion radius $r_\Ta$,  using the definitions Eq.\il(\ref{Eq:lLE}).
As seen in Sec.\il(\ref{Sec:counter-rotating-inversion-NSflow}), a counter-rotating inversion point \emph{must} be  out of the ergoregion, as confirmed also  from the analysis  in Figs\il(\ref{Fig:Plotsigmamax},\ref{Fig:Plotsigmalimit},\ref{Fig:Plotvinthallo},\ref{Fig:Plotvinthalloa}).
 Therefore, for  $\ell<0$ ($\ell=\ell^+<0$ or $\ell=\ell^-<0$, defined in  $a\in\,]\,1,a_2\,[$ and $r\in\,]\,r_0^-,r_0^+\,[\,-\,]\,r_\delta^-,r_\delta^+\,[$) there are \emph{no} inversion points  in  $[\,r_\epsilon^-,r_\epsilon^+\,]$.

 There are inversion points $\sigma=\sigma_\Ta\in\,[\,0,1\,]$  in:
 \bea&&
 \ell <0, r_\Ta\in\,]\,0, r_b\,]\quad \mbox{or, more precisely}
\quad
(\ell<0, r_\Ta>0)\quad \mbox{and}\quad \left(\ell\in\,\left[\,\ell_\beta,0\,\right[\, , r_\Ta\leq 2\right).
 \eea
Focusing on the  ergoregion,
 there are inversion points  $\sigma=\sigma_\Ta\in\,[\,0,1\,]$ for
{
 \bea&&
r_\Ta\in\,]\,0, r_{\epsilon}^-\,[\, :\quad  \ell \leq \ell_\rho<0,\quad  0<r_\Ta\leq r_b<1;
\\\nonumber
&&\quad r_\Ta>r_\epsilon^+:\quad r_\Ta\in\,[\,r_B^+,r_b\,]\, ,
\quad \mbox{where}\quad
 r_B^\mp\equiv\frac{1}{a \ell }+1-\frac{a}{\ell }\mp\sqrt{\frac{\left(a^2-1\right) \left(a^2-2 a \ell -1\right)}{a^2 \ell ^2}}.
\\&&\nonumber \mbox{more precisely:}\quad
 \left(r_\Ta\in\,]\,1,2\,]\, , \ell \leq\ell_\rho\right), \left(r_\Ta>2, \ell\in\,[\,\ell_\beta,\ell_\rho\,]\, \right).
 \eea}
{In {BH} case the region $]\,0,r_{\epsilon}^-\,[$ is not observable by the distant observer, constituting  a main difference with the super-spinars where  counter-rotating inversion points are possible in this region.}
 \subsubsection{Counter-rotation at the inversion point}\label{Sec:sever-amernothermed}
We explore   the conditions for $u^\phi=0$ and $\ell<0$,
 within $(\Em>0, \La<0)$ constant, using  definition Eq.\il(\ref{Eq:lLE}).  (In this analysis we do not consider the normalization condition.). This case includes tori with specific angular momentum $\ell=\ell^+<0$ and $\ell=\ell^-<0$ (possible for $a<a_2$ with $\Em>0$ and $\La<0$ in $r\in\,]\,r_0^-,r_0^+\,[\,-\,[\,r_\delta^-,r_\delta^+\,]$).

There is
\bea&&\label{Eq:imper-corp-specistat}
 (\ell\leq0):\quad u^\phi=0\;(u^t>0) \quad\{\Em\geq 0,  \La\leq0\},
\quad
 (\sigma_\Ta\in\, ]\,0,\sigma_\epsilon\,[\, , r_\Ta>0),  (\sigma_\Ta\in\, [\,\sigma_\epsilon,1\,]\, , r_\Ta\in\, ]\,0,r_\epsilon^-\,[\, \cup\, r_\Ta>r_\epsilon^+)
\eea
where the velocity $u^t$, the  energy $\Em$ and angular momentum $\La$ are
\bea\label{Eq:attr-qua-giap}
&&
u^t= \frac{-\La \Sigma_\Ta }{2 a r_\Ta \sigma_\Ta }= \frac{-\Em \Sigma_\Ta}{2 r_\Ta-\Sigma_\Ta},\quad \Em= \frac{2 r_\Ta+\Sigma_\Ta}{\Sigma_\Ta}u^t= \frac{-\La \left[\Sigma_\Ta-2 r_\Ta \right]}{2 a r_\Ta \sigma_\Ta } ,\quad \La=-\frac{2 a r_\Ta \sigma_\Ta  u^t}{\Sigma_\Ta}= -\frac{2 a \Em r_\Ta \sigma_\Ta }{\Sigma_\Ta-2 r_\Ta}.
\eea
The case  $\ell=0$  is possible  only for $\La=0$ and  $\sigma=0$ (on the {BH} axis) where
$
\Em={ u^t \Delta}/({a^2+r^2})
$, which is considered in Sec.\il(\ref{Sec:corot-inversion})
{Hence below we focus on  the   time-like and photon-like components.}

\medskip

\textbf{Time-like and photon-like components}

\medskip

For  $(\Em>0,\La<0)$ the analysis is in Eq.\il(\ref{Eq:imper-corp-specistat}). Further constraints  follow from the  normalization condition on the  flow components.
Let us define
:
 \bea&&
\Em_{\Gamma}\equiv\frac{\Sigma_{\Ta}-2 r_{\Ta}}{\Sigma_{\Ta}} \dot{t_{\Ta}},\quad\La_{\Gamma}=-\frac{2 a r_{\Ta} \sigma_{\Ta}  \dot{t}_{\Ta}}{\Sigma_{\Ta}},\quad \dot{\theta}^2_{\Gamma}\equiv\frac{\tilde{\mathbf{x}}_\Ta-a^2 (\sigma_{\Ta} -1) \left(\dot{t}_{\Ta}^2-1\right)}{\Sigma_{\Ta}^2},\quad  \tilde{\mathbf{x}}_\Ta\equiv r_{\Ta} \left[(r_{\Ta}-2) \dot{t}_{\Ta}^2-r_{\Ta}\right]
 \\\nonumber
 &&\dot{r}^2_{\Gamma}\equiv-\frac{\Delta_{\Ta} \left[a^4 (\sigma_{\Ta} -1)^2 \dot{\theta}_{\Ta}^2+a^2 (\sigma_{\Ta} -1) \left(\dot{t}_{\Ta}^2-1-2 r_{\Ta}^2 \dot{\theta}_{\Ta}^2\right)+r_{\Ta}^4 \dot{\theta}_{\Ta}^2-\mathbf{\tilde{x}}_\Ta\right]}{\Sigma_{\Ta}^2},
 \quad  \dot{t}_{\Gamma}\equiv\sqrt{\frac{\Sigma_{\Ta}}{\Sigma_{\Ta}-2r_{\Ta}}}.
 \eea
For time-like particles,  inversion points are for   $(\dot{r}_{\Ta}^2=\dot{r}^2_{\Gamma}, \La=\La_{\Gamma}, \Em=\Em_{\Gamma})$ and
 \bea&&
 \sigma_{\Ta}\in \,]\,0,\sigma_\epsilon\,[\, :\,(\dot{t}_{\Ta}=\dot{t}_{\Gamma}, \dot{\theta}^2=0); (\dot{t}_{\Ta}>\dot{t}_{\Gamma}, \dot{\theta}_{\Ta}^2\, \in\,[\,0, \dot{\theta}^2_{\Gamma}\,\,]\,);
 \\&&\nonumber
 \sigma_{\Ta} =\sigma_\epsilon: (r_{\Ta}\neq r_\epsilon^-, (\dot{t}_{\Ta}=\dot{t}_{\Gamma}, \dot{\theta}_{\Ta}^2=0); (\dot{t}_{\Ta}>\dot{t}_{\Gamma}, \dot{\theta}_{\Ta}^2\,\in\,[\,0, \dot{\theta}^2_{\Gamma}\,\,]\,));
 \\\nonumber
 &&\sigma_{\Ta}\in \,]\,\sigma_\epsilon,1\,]\,:\, (r_{\Ta}\in \,]\,0,r_\epsilon^-\,[\, \cup \, \, r_{\Ta}>r_{\epsilon}^+, (\dot{t}_{\Ta}=\dot{t}_{\Gamma}, \dot{\theta}_{\Ta}^2=0); (\dot{t}_{\Ta}>\dot{t}_{\Gamma}, \dot{\theta}_{\Ta}^2\in\, [\,0, \dot{\theta}^2_{\Gamma}\,\,]\, ).
 \eea
 Let us introduce:
 \bea&&
\dot{\theta}^2_X\equiv\frac{\dot{t}_{\Ta}^2 \left[\Sigma_{\Ta}-2r_{\Ta} \right]}{\Sigma_{\Ta}^2};\quad  \dot{r}^2_X=-\frac{\Delta_{\Ta} \left[a^4 (\sigma_{\Ta} -1)^2 \dot{\theta}_{\Ta}^2+a^2 (\sigma_{\Ta} -1) \left(\dot{t}_{\Ta}^2-2 r_{\Ta}^2 \dot{\theta}_{\Ta}^2\right)+r_{\Ta}^2 (r_{\Ta}^2 \dot{\theta}_{\Ta}^2-1)-\tilde{\mathbf{x}}_\Ta\right]}{\Sigma_{\Ta}^2}.
 \eea
For
photon--like particles there is  $(\dot{r}_{\Ta}^2=\dot{r}^2_X, \La=\La_{\Gamma}, \Em=\Em_{\Gamma})$, with
 \bea&&\nonumber
  	\sigma_{\Ta}\in \,\,]\,0,\sigma_\epsilon\,[\,: \dot{\theta}_{\Ta}^2\in\, [\,0 ,\dot{\theta}^2_X\,\,]\,;\,\quad\sigma_{\Ta} =\sigma_\epsilon: (r_{\Ta}\neq r_\epsilon^-\,[\, , \dot{\theta}_{\Ta}^2\,\in\, [\,0 ,\dot{\theta}^2_X\,\,]\,);
  \;\;  \sigma_{\Ta}\in \,\,]\,\sigma_\epsilon,1\,]\,: (r_{\Ta}\in \,]\,0,r_\epsilon^-\,[\,\cup \, r_{\Ta}>r_\epsilon^+,  \dot{\theta}_{\Ta}^2\in\, [\,0 ,\dot{\theta}^2_X\,\,]\,).
 \eea
 {Although the inversion spheres are similarly defined for photons and matter components by the  $\ell$--parameter range only, the inversion point coordinate and the velocity components  distinguish photon--like and  time-like components differently according to the super-spinars spin.}
\subsection{Inversion points of the co-rotating  ($\ell>0$) flows}\label{Sec:corot-inversion}

There are \emph{no} inversion points for $\ell>0$ with $\La\geq 0$ and $\Em\geq 0$.
A necessary condition for the existence of an inversion point, from definition Eq.\il(\ref{Eq:lLE}) of $\La$ at $(r_\Ta,\sigma_\Ta)$, is $\La<0$ and   therefore it occurs in the co-rotating case only for $\La<0$ and $\Em<0$.
(However  these solutions are only for spacelike (tachyonic) particles.).
\subsection{Inversion points: limiting case $\ell=0$}\label{Sec:inversionlequal0}
In the {NSs} with spin $a\in[1,a_2]$ there are orbits with angular momentum  $\La=0$,  therefore with specific angular momentum $\ell=0$, (and $\Em>0$) on $r_0^\pm$ with  $r_0^-\leq r_{0}^+<r_\epsilon^+=2$ (on the equatorial plane).  Within these conditions inversion points can  form  in some limiting cases
$ (\sigma =0, r\geq 0),(\sigma\in\,]\,0,1\, [\, ,  r=0)
$.
However within  condition $\ell=0$ there could be   $u^\phi\neq0$, i.e., ZAMOs are not (generally) zero angular velocity observers  ({ZAVOS}).

\section{Discussion and Conclusions}\label{Sec:discussion}
Inversion  coronas are spherical shells surrounding the central singularity,  interpreted   as a  property of the background geometry which can   identify  the central attractor,  distinguishing    {BHs} and {NSs}, and the  accretion configuration characteristics   defining accretion or  proto-jets driven  flows. Inversion coronas are generally  closed spherical shells  with  narrow thickness and small  extension on the equatorial plane and rotational central axis, varying  little with the   fluid initial conditions and the details  of the  emission processes, having   therefore  a  possible remarkable   observational significance  and
  applicability  to  various  orbiting toroidal  models.

We  focused particularly   on the estimation of the coronas  maximum extension  along  the central axis and on the equatorial plane, analysing   the coronas thickness, the  constraints  to the verticality and   equatorial extension of the inversion points corona,  for accretion driven and proto-jets driven flows.
The  inversion surface extension on  the central rotational axis, maximum vertical  height, is  of the order of $\lesssim 1.44$ for accretion driven flows.

  We studied function $\ell_\Ta$, defining the inversion surface in all generality,   later assuming constraints on the  constants   ($\Em$,$\La$) signs,  fixing the  normalization constant and  studying the inversion coronas for tori driven or proto-jets driven inversion points.
 We discussed   the definition of fluids and particles co-rotation and counter-rotation, in relation to the   $\{\ell, \La, \Em\}$  sign and   the {NS} casual structure.
 Co-rotating flows ($a\ell>0$) have no (timelike or photon like particle) inversion  points.

The disk inner region  is an extremely active part of the disk and, although the free falling hypothesis we consider  is widely used and justified, the flow in the  inner edge  region could  be exposed  to several different factors. The inner region of the disk can be
 subjected to  Pointing effects and  magnetic fields.
Inner edge can also be distorted by local oscillations affecting the -accretion rate. High-frequency quasi periodic oscillations
 could also modulate the accretion disk in oscillatory modes or  having  stabilizing effects.
  All these aspects can be of great importance for the study of the components of the accretion and emission flux of jets,  they are however  not considered in this first work on the analysis of azimuthal inversion points. On the other hand, the characteristics demonstrated for inversion  coronas, defined as background property, with little internal variation depending on the particle specific angular  momentum assure of their significance and very broad applicability.

{BHs} and super-spinning  geometries are also distinguished by  the presence of double inversion points for  planes different from the equatorial. This feature,  which can be inferred from Figs.\il(\ref{Fig:Plotvinthallo},\ref{Fig:Plotvinthalloa}), has profound consequences for  particles motion and the observational properties of the coronas. We argue that the inner corona region, closest to the central attractor, is the most active part  and observably recognizable.

  The results  of this  analysis therefore  individuate  strong  signatures of Kerr super-spinars,  in both accretion  and jet counter-rotating, flows,
  distinguishing Kerr {NSs} an {BHs}, and  a relevance of  our findings  is in the fact that, as for the {BH} case,  the inversion corona is a narrow region surrounding the central attractor, independent by the details of the accretion processes, the emission mechanisms  and tori models  which are  plagued, generally, by  large arbitrariness.  Inversion coronas  are  distinguished for proto-jets and accretion driven flows.

{Our findings   discern  {BHs}  and super-spinning objects, depending on   the central object spin  $a$    and the different orbiting objects, differentiating co-rotating  and counter--rotating tori and proto-jets, inner and outer tori of an orbiting couple.
The flow    is located in  restricted orbital range localized  in the orbital cocoon, inversion coronas, surrounding the central singularity, out of  ergoregions.
Inversion surfaces
are  ultimately a geometric property  of the Kerr spacetime  emerging, in the conditions analyzed here, for the free flows of particles  and photons changing on these surfaces the  toroidal component  $u^\phi$, and therefore  the surfaces could be observed from the effects related to the change in toroidal velocity, the condition $u^\phi=0$.
  While we recognize that  more work is required  to assert more than the possibility of observational relevance of our  results,
   we expect   that the inversion coronas   might   manifest as  an active  part of the accreting flux of matter and photons,   particularly the region  close  to the central singularity and the rotational axis  which could be  characterized  by an increase of the flow luminosity and temperature, given the particular structure of the surfaces in that region, see Fig.\il(\ref{Fig:Plotsigmalimit3W}) (or vice versa, the effects considered here could  distinguish the region from the emissions with a component subject to the inversion sphere by  emerging the peculiarity of the folding of the surface at the poles).
 On the other hand, the possible  observational properties  related to the flows in  this region    depend strongly on the  processes timescales, i.e. they depend on   the  time flow reaches the inversion points, reached at  different times $t_\Ta$ depending  on the initial data.}

\acknowledgements
The authors are grateful to RCTPA, IoP, Silesian University in Opava.

\appendix
{\section{Kerr Naked singularities spacetime  circular geodesic structure }\label{Appendix:radii}
For convenience we provide below an explicit expression for the set of radii  $\{r_{ms}^+,r_{mb}^\pm, r_\gamma^+,\tilde{r}_{ms}^-,\bar{r}_{ms}^-\}$ of the Kerr NS spacetime geodesic structure, introduced in Sec.\il(\ref{Sec:extended-geo-struc}).  Hence, there is
\bea&&
r_{ms}^+\equiv 3+ \upsilon ^++\Upsilon,\quad
r_{mb}^+\equiv 2+ a+2 \sqrt{a+1},\quad  r_\gamma^+\equiv 4 \cos ^2\left(\frac{1}{6} \cos ^{-1}\left[2 a^2-1\right]\right)
\\\nonumber
&&
\tilde{r}_{ms}^-\equiv 3-\upsilon ^--\Upsilon,\quad \bar{r}_{ms}^-\equiv\upsilon ^--\Upsilon +3,\quad  r_{mb}^-\equiv 2+ a-2 \sqrt{a+1},
\eea
where
\bea
\upsilon ^{\mp }\equiv \sqrt{B [(H\mp 2 \Upsilon )+4]},\quad \Upsilon \equiv \sqrt{3 a^2+(H+1)^2},\quad B\equiv 2-H,\quad H\equiv \sqrt[3]{1-a^2} \left(\sqrt[3]{1-a}+\sqrt[3]{a+1}\right)
\eea
(see Figs\il(\ref{Fig:Plotdicsprosc})).
We  introduce  also radii  $\{r_0^\pm,r_{\delta}^\pm\}$ as:
\bea&&
r_0^\mp \equiv \frac{\sqrt{\hat{K}}\mp X}{\sqrt{6}},\quad
r_\delta^-\equiv \frac{8}{3} \sin ^2\left[\frac{1}{6} \cos ^{-1}\left(1-\frac{27 a^2}{16}\right)\right],\quad r_\delta^+\equiv \frac{4}{3} \left[\sin \left(\frac{1}{3} \sin ^{-1}\left[1-\frac{27 a^2}{16}\right]\right)+1\right],
\eea
where
\bea
 X\equiv \sqrt{a^2 \left(\frac{6 \sqrt{6}}{\sqrt{\hat{K}}}-6\right)-\hat{K}},\quad  \hat{K}\equiv \frac{4 a^4}{\sqrt[3]{T}}-2 a^2+\sqrt[3]{T},\quad
T\equiv 3 a^4 \left[\sqrt{81-48 a^2}+9\right]-8 a^6.
\eea
}
%\appendix
%\input{appendixI-a}

\end{document}